\documentclass[floats,floatfix,showpacs,amssymb, amsmath,prd,twocolumn,superscriptaddress,nofootinbib,nolongbibliography,reprint]{revtex4-2}

\usepackage[T1]{fontenc}
\usepackage{lmodern}
\usepackage{amssymb,amsmath,verbatim,mathtools,needspace,enumitem,etoolbox,graphicx,physics,microtype,afterpage,bigints,gensymb,tabularx,xspace} %
\usepackage{journals}
\usepackage{siunitx,seqsplit} 

\usepackage{float}
\usepackage{tikz}
\usetikzlibrary{fadings}
\usetikzlibrary{calc}
\usetikzlibrary{arrows.meta,positioning,fit,matrix}
\usetikzlibrary{decorations.pathreplacing}

\usepackage{xcolor}
\definecolor{linkcolor}{rgb}{0.0,0.3,0.5}
\definecolor{dodgerblue}{HTML}{1E90FF}
\usepackage[unicode, colorlinks=true, linkcolor=linkcolor, citecolor=linkcolor, filecolor=linkcolor,urlcolor=linkcolor, pdfusetitle]{hyperref}
\usepackage[all]{hypcap}

\hypersetup{
    breaklinks=true   
}

\usepackage[utf8]{inputenc}

\usepackage{orcidlink}
\usepackage{algorithm}
\usepackage{algorithmic}
\usepackage[normalem]{ulem}

\renewcommand{\vec}[1]{\boldsymbol{#1}}

\makeatletter
\newcommand*{\balancecolsandclearpage}{\close@column@grid \cleardoublepage \twocolumngrid}
\makeatother

\newcommand{\milan}{\affiliation{Dipartimento di Fisica ``G. Occhialini'', Universit\'a degli Studi di Milano-Bicocca, Piazza della Scienza 3, 20126 Milano, Italy}}
\newcommand{\infn}{\affiliation{INFN, Sezione di Milano-Bicocca, Piazza della Scienza 3, 20126 Milano, Italy}}
\newcommand{\AEi}{\affiliation{Max Planck Institute for Gravitational Physics (Albert Einstein Institute) Am Mühlenberg 1, 14476 Potsdam, Germany}}
\newcommand{\thessaloniki}{\affiliation{Department of Physics, Aristotle University of Thessaloniki, Thessaloniki 54124, Greece}}
\newcommand{\noa}{\affiliation{Institute for Astronomy, Astrophysics, Space Applications and Remote Sensing, National Observatory of Athens, 15236 Penteli, Greece}}

\newcommand{\review}[1]{{\color{black}#1}}

\newcommand{\f}{f}
\newcommand{\fdot}{\dot{f}}
\newcommand{\rescaledamp}{\tilde{\mathcal{A}}}
\newcommand{\rescaledfdot}{\tilde{\dot{f}}}
\newcommand{\amp}{\mathcal{A}}
\newcommand{\tobs}{T_\text{obs}}
\newcommand{\ppop}{p_\text{pop}}
\newcommand{\sinstr}{S_\text{instr}(f)}
\newcommand{\sconf}{S_n(f)}
\newcommand{\boldsconf}{\vec{S}_n}
\newcommand{\avgsconf}{S_n(f)}

\newcommand{\sgal}{S_\text{gal}(f)}
\newcommand{\AET}{\text{AET}}
\newcommand{\corr}{\varrho}
\newcommand{\alphaPL}{\alpha_\text{PL}}
\newcommand{\alphaGamma}{\alpha_\Gamma}
\newcommand{\betaGamma}{\beta_\Gamma}
\newcommand{\muGamma}{\mu_\Gamma}
\newcommand{\Nbinaries}{N_\text{b}}
\newcommand{\mugpr}{\mu_\text{\tiny GPR}}
\newcommand{\sigmagpr}{\sigma_\text{\tiny GPR}}
\newcommand{\pp}{$p$--$p$\xspace}

\usepackage{array}
\usepackage{lipsum}
\usepackage{soul}

\begin{document}

\title{Inferring the population properties of galactic binaries\\ from LISA’s stochastic foreground}

\author{Federico De Santi$\,$\orcidlink{0009-0000-2445-5729}}
\email{f.desanti@campus.unimib.it}
\milan \infn

\author{Alessandro Santini$\,$\orcidlink{0000-0001-6936-8581}}
\AEi

\author{Alexandre Toubiana$\,$\orcidlink{0000-0002-2685-1538}}

\milan \infn

\author{\\Nikolaos Karnesis$\,$\orcidlink{0000-0002-2380-3186}}
\thessaloniki \noa

\author{Davide Gerosa$\,$\orcidlink{0000-0002-0933-3579}}

\milan \infn

\begin{abstract}
Galactic binaries are expected to be the most numerous LISA sources and to produce a stochastic gravitational-wave foreground whose spectral shape encodes information about the underlying population. 
Extracting this information with standard hierarchical methods is challenging due to the high dimensionality of the problem 
and the computational cost of global-fit analyses.
We present a simulation-based inference framework to measure the population properties of galactic binaries directly from the 
reconstructed foreground. Adopting an astrophysically agnostic parametrization in the observable space—defined by signal amplitude, frequency, and frequency derivative—we generate synthetic catalogs and foreground spectra using a global-fit-inspired subtraction algorithm. We then train a neural posterior estimator to map spectra to population parameters.
We validate our method on simulated data and recover population parameters with good accuracy, including the total number of binaries. 
As a by-product, we present a GPU-accelerated version of the subtraction algorithm, which delivers a $\sim 10^2\times$ speed-up compared to previous implementations in the literature. 
Our results demonstrate that LISA’s stochastic foreground alone carries significant information about the Galactic binary population and provide a practical step toward joint inference from resolved and unresolved sources.
\end{abstract}

\maketitle

\section{Introduction}
The Laser Interferometer Space Antenna (LISA) is a space-based gravitational-wave observatory designed
to detect gravitational waves (GWs) in the frequency band from $10^{-4}\,\mathrm{Hz}$ to $10^{-1}\,\mathrm{Hz}$~\cite{LISA:2024hlh}.
Among the targeted astrophysical sources, galactic binaries (GBs), mostly double white dwarfs (DWDs),
are expected to be the most abundant, with the Milky Way and its satellites containing up to $\mathcal{O}(10^{7})$
such systems~\cite{Lamberts:2019nyk,Korol:2020hay,Korol:2021pun,2024A&A...692A.165T, Nelemans:2001hp,Ruiter:2007xx,Breivik:2017jip}.
Their continuous GW emission will form a confusion noise, usually referred to as the Galactic foreground.
Around $1\,\mathrm{mHz}$, this Galactic foreground is expected to dominate over the instrumental noise in the total noise budget when analysing LISA data.
In addition, about $10^{4}$ GBs are expected to be individually resolved.
LISA data analysis will be performed by inferring the properties of resolvable sources, as well as their number and the total noise level, via a global fit~\cite{Littenberg:2020bxy, Littenberg:2023xpl, Katz:2024oqg, Strub:2024kbe, Deng:2025wgk}.

The Galactic foreground is, however, not only an additional source of noise for the analysis,
but also a valuable source of information about the GB population. Recent work has explored population
inference focusing just on resolved binaries, e.g.~\cite{Delfavero:2024zyl}, while Ref.~\cite{Georgousi:2022uyt} has linked elemental population properties, 
surch as the overall mass, to the shape of the confusion noise. 
Ultimately, astrophysical inference on the GB population from the global fit output will require combining
information from the Galactic foreground and from the resolved sources in a self-consistent manner~\cite{2026arXiv260104168T}. A key ingredient for this is establishing the relation between
the Galactic foreground and the GB population, which is the goal of this paper.
For completeness, we also mention a complementary simulation-based approach that operates on realizations
of the LISA datastream~\cite{Srinivasan:2025etu}, without relying on global fits to perform astrophysical inference.

We adopt a simulation-based inference (SBI) framework~\cite{2020PNAS..11730055C,2025arXiv250812939D}
to infer a posterior distribution for the population parameters $\vec{\Lambda}$ given the observed confusion noise $\avgsconf$.
SBI methods do not require an explicit likelihood function, as they rely on forward simulations to learn the mapping between parameters and data.
This approach is particularly well suited to our case, where the probability $p(\avgsconf \mid \vec{\Lambda})$
is not analytically tractable.
A similar inference paradigm has already been applied in Ref.~\cite{Alvey:2023npw}
to the inference of stochastic backgrounds in LISA, though focusing also on cosmological signals rather than astrophysical ones.

We simulate foregrounds using the subtraction algorithm first presented in Ref.~\cite{Karnesis:2021tsh}, which is explicitly designed to mimic an idealised global fit-like process.
The algorithm progressively identifies resolvable sources in the data, thus providing an optimistic but also realistic estimate of the remaining foreground.
Relative to the original implementation~\cite{Karnesis:2021tsh}, we optimized the code by exploiting GPU parallelization, leading to
a speed-up of a factor of $\sim10^{2}$ in both data generation and source subtraction.
This enables the processing of a catalog with $10^{7}$ sources in $\mathcal{O}(1\,\mathrm{min})$.

We generate mock catalogs using astrophysically agnostic population models. In particular, we directly parametrize
the GB population in terms of observable quantities, namely the binary frequency, its time derivative, and the signal amplitude.
Working directly in this observable space reduces uncertainties associated with astrophysical modeling while retaining the flexibility
to capture a wide range of astrophysical scenarios.
Our parametrization is inspired by existing astrophysical population models and is able to accurately reproduce their predictions,
as demonstrated by comparison with the catalog of Ref.~\cite{Lamberts:2019nyk}.

We test our method on a suite of simulated foregrounds. For these tests, we further employ a Markov chain Monte Carlo (MCMC) reconstruction of the
power spectral density (PSD)~\cite{Santini:2025iuj} to mimic a realistic inference from a global fit output, in which the foreground is not directly observed but instead inferred.
We successfully recover the parameters of the injected foregrounds, including the total number of sources.
We find that the parameters governing the frequency and amplitude distributions are better constrained than others,
indicating that certain aspects of the population have a more pronounced impact on the shape of the foreground.
When applying our method to the simulations of Ref.~\cite{Lamberts:2019nyk},
we identify and address technical challenges that must be overcome when relating the foreground inferred from a global fit to the underlying astrophysical population.

This paper is organized as follows. In Sec.~\ref{sec:GBs}, we  
describe the parametrization for the population and the strategy adopted to generate mock catalogs. In Sec.~\ref{sec: SBI}, we present our SBI framework. 
In Sec.~\ref{sec: results}, we present the results of our analysis detailing 
 performance and validation of our method. 
Finally, in Sec.~\ref{sec: conclusions}, we summarize our findings
and discuss their implications.

\section{Galactic Binaries}\label{sec:GBs}
\subsection{Signal parametrization}\label{sec:gwemission}

GW emission from GBs is characterized by the parameters
\begin{equation}
\vec{\theta} = \{\amp, \f, \fdot, \iota, \phi_0, \psi, \lambda, \beta\}\,,
\end{equation}
namely the amplitude $\amp$, frequency $\f$, frequency derivative $\fdot$, inclination $\iota$, initial phase $\phi_0$,
polarization $\psi$, ecliptic longitude $\lambda$, and ecliptic latitude $\beta$.
We restrict ourselves to GBs on quasi-circular orbits, as systems with significant eccentricity are expected to be extremely rare
($\lesssim 0.1\,\%$ of the total)~\cite{2025A&A...702A.131H,2025A&A...704A.156R}.
The GW signal emitted by the source is
\begin{equation}\label{eq: GB waveform}
    \begin{split}
        h_+(t) &= \frac{1}{2}\,\amp \left( 1 + \cos^2\iota \right) \cos\phi(t)\,,\\
        h_\times(t) &= \amp \cos\iota \sin\phi(t)\,,
    \end{split}
\end{equation}
where the phase $\phi(t)$ can be Taylor-expanded to first order, i.e.
\begin{equation}\label{eq: phase Taylor expansion}
    \phi(t) \simeq \phi_0 + 2\pi \f t + \pi \fdot t^2\,.
\end{equation}
The ecliptic coordinates specify the sky position of the source in the solar system barycenter frame and, together with the polarization angle, enter the LISA response function~\cite{Cornish:2007if}.

Under the assumption that the binary evolution is entirely driven by energy loss due to GW emission,
one has $\fdot=\fdot_\text{gw}$ with
    \begin{align}
        \fdot_\text{gw} &= \frac{96}{5} \pi^{8/3} \mathcal{M}^{5/3} \f ^{11/3}\label{eq: fdot gw} 
        \end{align}
        and 
        \begin{align}
        \amp & = \frac{2\mathcal{M}^{5/3}\pi^{2/3}}{d_L}\f^{2/3}\label{eq: amplitude gw}\,,
    \end{align}
 where $\mathcal{M} = {(m_1m_2)^{3/5}}{(m_1+m_2)^{-1/5}}$ is the chirp mass of the binary and  $d_L$ is the luminosity distance.

In general, the evolution of GBs is influenced by a variety of astrophysical processes, like matter interactions, that can significantly alter 
their behavior, thus leading to deviations from the above relations, i.e.
\begin{equation}\label{eq: fdot astrophysical}
  \fdot = \fdot_\text{gw} + \fdot_\text{astro}\,.
\end{equation}
In the case of DWDs, for instance, tidal torques and mass transfer are the main contributors to $\fdot_\text{astro}$. Tidal torques act early in the evolution 
and drive spin-orbit synchronization, typically adding a positive contribution to $\fdot$ ~\cite{2015ApJ...806...76K,1977A&A....57..383Z, 2002MNRAS.329..897H}. If the separation of the binaries becomes sufficiently small, the less massive star overfills its Roche lobe and starts transferring mass to its companion. Mass transfer can lead to two qualitatively different outcomes: if unstable, it drives a rapid
coalescence, hence reducing the number of detectable sources, while if stable can produce long-lived interacting systems (AM~CVn-like binaries~\cite{2010PASP..122.1133S}) that outspiral~\cite{2010PASP..122.1133S,2015ApJ...805L...6S,
Maoz:2013hna,Shen:2017flp,2024A&A...692A.165T}, hence $\fdot<0$. Incorporating these effects into a semi-analytic evolutionary model applied to a simulated population of DWDs at formation, 
Ref.~\cite{2024A&A...692A.165T} showed that the resulting foreground can differ significantly from 
that obtained under GW-driven evolution alone.

State-of-the-art GB catalogs are produced with population-synthesis codes~\cite{Korol:2017qcx,Lamberts:2019nyk,Korol:2021pun, 2026arXiv260211765M},
based on cosmological simulations of the Milky Way and prescriptions of stellar evolution. 
In the GW-only scenario, the conversion from ($\mathcal{M}, d_L$) to the observables ($\fdot, \amp$) is through Eqs.~\eqref{eq: fdot gw} and \eqref{eq: amplitude gw}. 
For interacting systems, a common approach is to couple synthesis outputs with semi-analytical evolutionary models ~\cite{2004MNRAS.350..113M,2015ApJ...806...76K,2021ApJ...908....1S,2024A&A...692A.165T},
which are physically motivated but inevitably rely on model-dependent assumptions.

Since the observable space ($f, \fdot, \amp$) is the one actually probed by LISA and enters in the waveform model, 
we instead adopt a complementary strategy to directly model the population in this space, without relying on specific astrophysical assumptions.
This allows for a flexible parametrization able to describe a broad range of physical scenarios 
without committing to a specific stellar evolution prescription.

\begin{table*}
  \centering
  \setlength{\tabcolsep}{12pt}  
  \renewcommand{\arraystretch}{1.2}
  \begin{tabular}{cc>{\hspace{5pt}}c}
    \hline\\[-2ex]
    \multicolumn{1}{c}{\text{}} &
    \multicolumn{1}{c}{\text{Population distribution}} &
    \multicolumn{1}{c}{\text{Prior range}} \\[1ex]
    \hline \hline
    \noalign{\vskip 10pt}   

    \text{Power law} & 
    $\displaystyle p(f\mid \alphaPL)
      \propto
      \begin{cases}
        f^{\alphaPL}, & 0.1\,{\rm mHz}\leq f \leq 10\,{\rm mHz},\\
        0, & \text{otherwise}
      \end{cases}$
    &
    \begin{tabular}[c]{@{}l@{}}
      $\alphaPL \sim \mathcal{U}(-4, -2)$\\
    \end{tabular}
    \\[5ex]
    \text{Gamma} & 
      $\displaystyle
      \begin{gathered}
      p(\rescaledfdot \mid \alphaGamma, \betaGamma)
        = \frac{\betaGamma^{\alphaGamma}}{\Gamma(\alphaGamma)}
          \,\rescaledfdot^{\,\alphaGamma-1} e^{-\betaGamma \rescaledfdot}\\
      \end{gathered}
      $
      &
      \begin{tabular}[c]{@{}l@{}}
        $\;\;\:\,\alphaGamma,\betaGamma \sim \mathcal{U}(1, 3)$\\
      \end{tabular}
      \\[3ex]
    \noalign{\vskip 5pt}   
    \text{Log-normal} & 
    $\displaystyle p(\rescaledamp \mid \sigma)
      =\frac{1}{\rescaledamp\,\sqrt{2\pi}\,\sigma}
       \exp\!\Big[-\frac{(\ln \rescaledamp)^2}{2\sigma^2}\Big]$
    &
    \begin{tabular}[c]{@{}l@{}}
      $\quad\sigma \sim \mathcal{U}(0.1, 0.3)$
    \end{tabular}
    \\[3ex]
    \hline
  \end{tabular}
  \caption{Building blocks of the population distribution $\ppop(\vec{\theta}\mid \vec{\Lambda})$ used in this work.
The corresponding prior distributions are indicated in the right column; the remaining priors are
$\muGamma \sim \mathcal{U}(-22,-20)$, $\mu \sim \mathcal{U}(-23, -21)$, $\corr \sim \mathcal{U}(0,1)$, and
$\Nbinaries \sim \mathcal{U}(5 \times 10^{6},\,30 \times 10^{6})$.
  }
  \label{tab: priors}
\end{table*}

\begin{figure}
  \centering
  \includegraphics[width=\columnwidth]{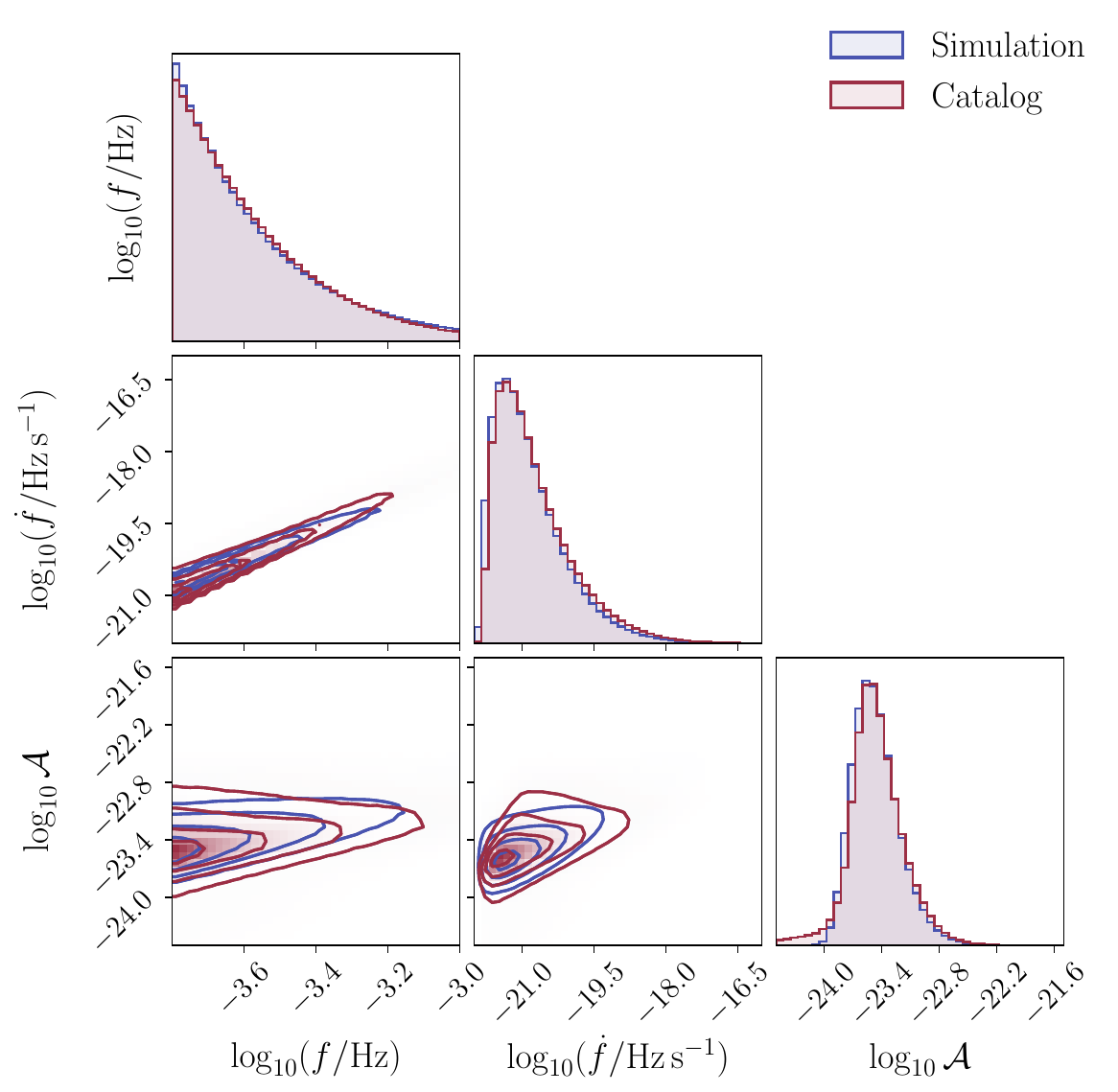}
  \caption{
  Comparison between the astrophysical catalog of Ref.~\cite{Lamberts:2019nyk} (red) and 
  a simulated catalog (blue) generated by our population model assuming 
  $\vec{\Lambda} = \{\alphaPL = -2.9, \alphaGamma = 1.96, \muGamma = -21, 
\betaGamma = 1.8, \mu = -22, \sigma = 0.18, \corr = 0.99, \Nbinaries = 7.5\times10^6\}$.
}
  \label{fig: astro cat vs simulation comparison}
\end{figure}

\subsection{Population model}
\label{sec: population model}

We model the GB population distribution $\ppop(\vec{\theta} \mid \vec{\Lambda})$, where
$\vec{\theta} = \{\f, \fdot, \amp\}$ are the observables and $\vec{\Lambda}$ are the population parameters.
In particular, we construct this parametrization with the goal of capturing
the main features of existing state-of-the-art catalogs while remaining as agnostic as possible regarding
the underlying astrophysics of binary evolution.
Specifically, we do not assume Eqs.~\eqref{eq: fdot gw},~\eqref{eq: amplitude gw} and~\eqref{eq: fdot astrophysical}
to relate $\fdot$ and $\amp$ to $\f$.

Our modeling choices are summarized in Table~\ref{tab: priors} and are guided by Ref.~\cite{Lamberts:2019nyk},
which provides a realistic astrophysical catalog of DWDs in a Milky Way-like galaxy, obtained by combining binary population synthesis
with cosmological simulations~\cite{2018MNRAS.480.2704L}.
This catalog, shown in red in Fig.~\ref{fig: astro cat vs simulation comparison}, includes
approximately $7.5$ million DWDs evolved under the assumption of purely GW-driven evolution.
Throughout the rest of this work, we focus on DWD-like populations with $\dot{f}>0$, 
but stress that our approach can be easily generalized to more complex astrophysical scenarios, including multiple
kinds of GBs. 

The marginal distribution of $f$ exhibits a decaying behavior that can be reasonably described by a power law.
This assumption follows from simple orbital scalings: 
the orbital period scales with the separation as $P \propto a^{3/2}$~\cite{Korol:2021pun}, 
whereas the separation itself is also expected to follow a power law distribution~\cite{Maoz:2012aj}.
Observations of binary white dwarfs from the Sloan Digital Sky Survey~\cite{2012ApJ...749L..11B} 
and the ESO-VLT Supernova-Ia Progenitor Survey support this behavior~\cite{Maoz:2016bxg, Maoz:2018epf}.

Inspecting the marginal distribution of $\log \fdot$ reveals that the rescaled quantity 
\begin{equation}
\label{rescaledfdot}
\rescaledfdot = \log_{10}\left(\frac{\fdot}{{\rm Hz \,s^{-1}}}\right) - \muGamma \, ,
\end{equation}
closely resembles a Gamma distribution, characterized by a concentration parameter $\alphaGamma$
and a rate parameter $\betaGamma$. 

The strong correlation between amplitude and frequency, which arises from Eq.~\eqref{eq: amplitude gw}, shown in Fig.~\ref{fig: astro cat vs simulation comparison} can be absorbed by defining
the rescaled logarithmic amplitude 
\begin{equation}\label{eq: rescaled amplitude gw}
    \rescaledamp =\log_{10} \left[ \left(\frac{f}{{\rm Hz}}\right)^{-2/3} \amp \right] -\mu\, .
\end{equation}
We then model the rescaled amplitude $\rescaledamp$ as a Log-normal distribution characterized by a standard deviation $\sigma$.

\begin{figure*}
  \centering 
  \hspace{-1cm}
  \includegraphics[width=0.9\textwidth]{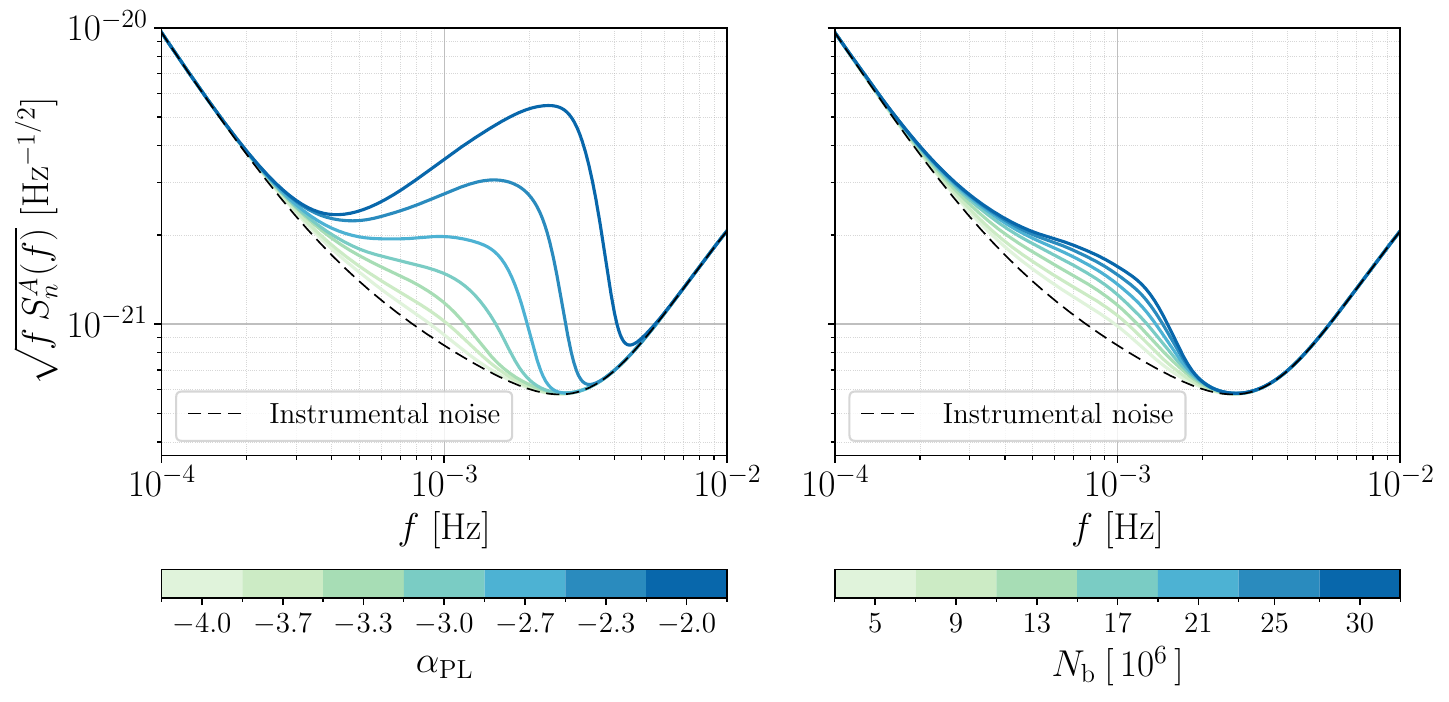}
  \caption{Characteristic strain sensitivity as a function of selected population parameters.
We show the predicted mean for the A channel assuming $\alphaGamma=2.5$, $\betaGamma=2$, $\muGamma=-20.5$, $\mu=-22$, $\sigma=0.3$, and $\varrho=0.85$.
In the left panel, we vary the spectral index of the frequency distribution $\alphaPL$ while fixing $N_{\rm b}=5\times10^{6}$.
In the right panel, we vary the total number of binaries $N_{\rm b}$ while fixing $\alphaPL=-3.8$.
  }
  \label{fig: gpr predictions}
\end{figure*}

Similar rescaling techniques do not appear to perform well in modeling the correlation between frequency and frequency derivative.
We therefore model the $f$--$\fdot$ correlation using a Gaussian copula~\cite{Adamcewicz:2022hce},
which depends on a correlation coefficient $\corr \in [0,1)$
(see Appendix~\ref{sec: copulas} for details).
With this choice, the resulting model can accommodate a wide range of correlations,
thereby allowing for deviations from purely GW-driven evolution.

Overall, our population model depends on eight dimensionless parameters
\begin{equation}\label{eq: pop hyperparameters}
  {\vec{\Lambda} = \{\alphaPL, \,\alphaGamma, \,\betaGamma, \,\muGamma, \,\mu, \,\sigma, \,\corr, \,N_\text{b}\}}\,,
\end{equation}
where the first seven population parameters characterize the population distribution as described above, and $N_{\rm b}$ is the overall number of binaries, which also needs to be inferred from the data.
Sampling the joint distribution $\ppop(\vec{\theta}\mid\vec{\Lambda})$
requires drawing samples from the marginal distributions of $\f$, $\rescaledfdot$, and $\tilde{\amp}$, rescaling back to $\fdot$ and $\amp$, and applying the copula to construct $p(f,\fdot)$.
The prior $p(\vec{\Lambda})$ is defined as a product of independent uniform priors with ranges
listed in Table~\ref{tab: priors}.
For $\Nbinaries$, we assume a uniform prior in the range $[5,30]\times 10^{6}$ in order to avoid unbalanced datasets in the subsequent inference.
This choice is motivated by current astrophysical predictions, e.g.\ Refs.~\cite{Korol:2021pun,Lamberts:2019nyk}.

Returning to Fig.~\ref{fig: astro cat vs simulation comparison}, we overplot in blue
the distribution obtained with our population model for
$\vec{\Lambda} = \{\alphaPL = -2.9, \alphaGamma = 1.96, \muGamma = -21, 
\betaGamma = 1.8, \mu = -22, \sigma = 0.18, \corr = 0.99, \Nbinaries = 7.5\times10^6\}$. 
This demonstrates that the parametrization described above is able to capture the main features of the astrophysical catalog,
both in terms of marginal distributions and correlations.
We stress that this comparison serves only as a consistency check, as no fitting procedure has been performed at this stage.

When simulating binaries, we further adopt standard priors for the extrinsic parameters, e.g.\ isotropic distributions for
the inclination $\iota$, phase $\phi_0$, and polarization $\psi$.
The sky position is directly sampled from the distribution of Ref.~\cite{Lamberts:2019nyk},
which models a Milky Way-like galaxy and its satellites as in Ref.~\cite{Wetzel:2016wro}.
As long as binaries with the same frequency are not placed at the same sky location, this choice does not impact the shape of the foreground.
We verified this by randomly re-sampling the sky locations and comparing the resulting outputs. 
Implicitly, we neglect the possibility that different regions of the Milky Way may host distinct populations. 
This could be straightforwardly incorporated by introducing multiple population components, which we leave for future work.
 
We conclude this section by briefly reviewing how the shape of the foreground is affected by some of the population
parameters. Figure~\ref{fig: gpr predictions} shows the characteristic strain\footnote{With little abuse of
notation, here $\sconf$ denotes the total PSD rescaled by the detector response (see Appendix~\ref{sec: from PSD to characteristic strain}). 
}
$\sqrt{f\,\sconf}$ of a set of simulated foregrounds obtained
by varying the power-law index $\alphaPL$ and the number of binaries $\Nbinaries$, as these parameters are intuitively expected to have a significant impact
on $\sconf$. We refer to Sec.~\ref{sec: SBI} for details on how these simulations were carried out.
For larger values of $\alphaPL$ (i.e.\ flatter distributions in frequency), both the amplitude and the peak frequency of the ``bump'' increase due to a higher
concentration of sources at high frequencies. We further observe an increase in the steepness of its high-frequency drop.
As further expected, the amplitude of the foreground increases with the number of binaries $N_{\rm b}$, cf.\ Ref.~\cite{Georgousi:2022uyt}.

\subsection{Power spectral density}
\label{psd}

The total PSD is the sum of the instrumental noise and the galactic foreground, i.e.
\begin{equation}\label{eq: confusion noise}
    S_n(f, \vec{\Lambda}) = S_\text{instr}(f) + S_\text{gal}(f, \vec{\Lambda})\,.
    \end{equation}
 We fix the instrumental noise $\sinstr$ to the
nominal \textsc{SciRDv1} LISA sensitivity curve from Ref.~\cite{LISA_SciRDv1}; details are reported in Appendix \ref{sec: lisa noise}.\\
The galactic foreground $\sgal$ is instead generated with the method described in Sec.~\ref{sec: data generation}, 
which mimics the process of a global fit and therefore provides a realistic estimate of the residual foreground after 
the subtraction of the resolved sources.

The foreground PSD is time-dependent: the longer the observation, the more sources are expected to be 
individually resolved~\cite{Karnesis:2021tsh,Buscicchio:2025zeb}. 
The observation time $\tobs$ is therefore a relevant hyperparameter. 
In this work, we focus on the result assuming the nominal $\tobs = 4 \mathrm{yr}$ for the LISA mission.
We also do not consider the yearly modulations induced by the spacecraft motion around the Sun, which make the foreground cyclo-stationary 
and dependent on the spatial distribution of sources~\cite{Pozzoli:2024wfe}.

\section{Simulation-based inference}\label{sec: SBI}

\tikzfading[name=bluefadeearly,
  left color=transparent!0,
  right color=transparent!100
]

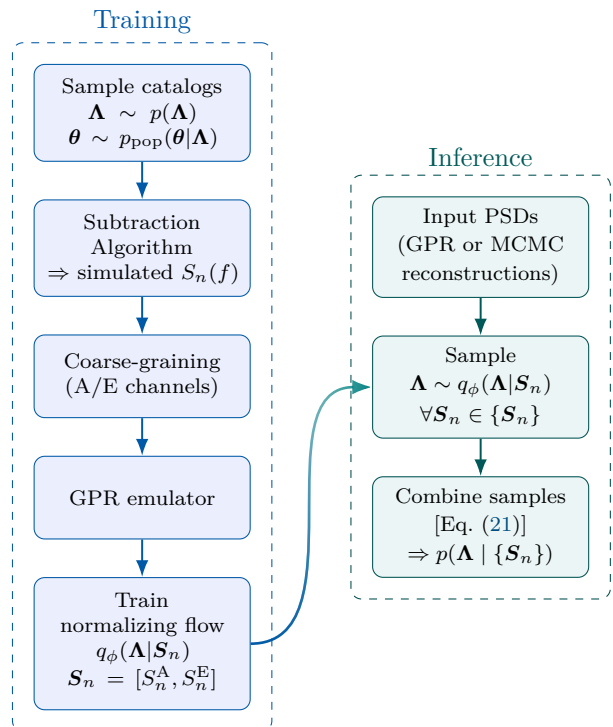
\begin{figure}[!t]
  \centering
  \begin{tikzpicture}[
    font=\footnotesize,
    >=Latex,
    node distance=2mm and 6mm,
    block/.style={draw, rounded corners, align=center, minimum height=11mm, text width=0.32\columnwidth, inner sep=1.5pt},
    aux/.style={draw, rounded corners, align=center, text width=0.3\columnwidth, inner sep=2.5pt},
    trainblock/.style={block, draw=blue!30!teal, fill=blue!70!teal!8},
    inferblock/.style={block, draw=teal!65!black, fill=teal!8},
    latentblock/.style={aux, draw=orange!75!black, fill=orange!10},
    trainline/.style={->, line width=0.75pt, draw=blue!30!teal},
    inferline/.style={->, line width=0.75pt, draw=teal!70!black},
    paneltrain/.style={draw=blue!30!teal, dashed, rounded corners, inner sep=8pt},
    panelinfer/.style={draw=teal!60!black, dashed, rounded corners, inner sep=8pt}
  ]

    \matrix (traincol) [
      matrix of nodes,
      nodes={trainblock},
      row sep=5mm,
      column sep=0mm
    ] {
      |[name=t1]| {\\Sample catalogs\\$\vec{\Lambda}\sim p(\vec{\Lambda})$\\$\vec{\theta}\sim \ppop(\vec{\theta}|\vec{\Lambda})$\smallskip} \\
      |[name=t2]| {\\Subtraction Algorithm\\$\Rightarrow$ simulated $\avgsconf$\smallskip} \\
      |[name=t3]| {Coarse-graining (A/E channels)} \\
      |[name=t4]| {GPR emulator} \\
      |[name=t5]| {\\Train\\ normalizing flow\\$q_\phi(\vec{\Lambda}| \boldsconf)$\\$\boldsconf = [S_n^\text{A}, S_n^\text{E}]$\medskip} \\
    };

    \draw[trainline] (t1) -- (t2);
    \draw[trainline] (t2) -- (t3);
    \draw[trainline] (t3) -- (t4);
    \draw[trainline] (t4) -- (t5);

    \node[paneltrain, fit=(t1)(t5),
      label={[text=blue!30!teal, font=\normalsize]above:Training}
    ] {};


    \path (t1.north) -- (t5.south) coordinate[midway] (trainCenter);
    \matrix (infercol) [
      matrix of nodes,
      nodes={inferblock},
      row sep=5mm,
      right=45mm of trainCenter,
      anchor=center
    ] {
      |[name=i1]| {\shortstack{\\Input PSDs\\(GPR or MCMC\\reconstructions)\smallskip}} \\
      |[name=i4]| {\shortstack{\\Sample\\ $\vec{\Lambda}\sim q_\phi(\vec{\Lambda}|\boldsconf)$\\$\forall \boldsconf \in \{\boldsconf\}$\smallskip}} \\
      |[name=i5]| {\shortstack{\\Combine samples\\$[$Eq.~\eqref{eq: posterior samples combination}$]$\\$\Rightarrow p(\vec{\Lambda}\mid\{\boldsconf\})$}\smallskip} \\
    };

    \draw[inferline] (i1) -- (i4);
    \draw[inferline] (i4) -- (i5);

    \node[panelinfer, fit=(i1)(i5),
      label={[text=teal!70!black, font=\normalsize]above:Inference}
    ] {};

  \draw[->,line width=1pt, draw=teal!100!green, path fading=bluefadeearly, fading angle=180]
    (t5.east) to[out=0,in=180,looseness=1.0] (i4.west);

  \draw[->, line width=1pt, draw=blue!30!teal, path fading=bluefadeearly, fading angle=90]
    (t5.east) to[out=0,in=180,looseness=1.0] (i4.west);

  \end{tikzpicture}
  \caption{Flowchart of our inference pipeline. The left column (blue) depicts the 
  data generation and processing in the training of the normalizing flow.
  The right column (teal) shows the inference phase, 
  where we perform inference over a set of PSD reconstructions $\{\boldsconf\}$,
  which can be obtained with different methods (e.g.\ GPR or MCMC).}
  \label{fig: sbi pipeline flowchart}
\end{figure}
\subsection{Simulations}\label{sec: data generation}

We adopt an SBI framework trained on reconstructing $p(\vec{\Lambda} | S_n)$, summarized in Fig.~\ref{fig: sbi pipeline flowchart}.
We simulate $\avgsconf$ using the subtraction strategy first proposed in Ref.~\cite{Karnesis:2021tsh} and here presented in Appendix \ref{sec: subtraction algorithm}. In short, the algorithm relies on an iterative procedure where the total PSD $\avgsconf$ is estimated by progressively removing the sources above a given signal-to-noise ratio (SNR) threshold.
The SNR is computed over the A, E, and T channels of time-delay interferometry (TDI)~\cite{Prince:2002hp,Hartwig:2021mzw}, i.e. 
\begin{equation}\label{eq: SNR inner product}
    \rho^2 = \sum_{i\in \{\text{A,E,T}\}} \langle h_i | h_i \rangle
\end{equation}
where  
\begin{equation}\label{eq: inner product}
    \langle h_i | h_j \rangle = 4 \Re \int_{f_\text{min}}^{f_\text{max}} \frac{\tilde{h}_i(f) \tilde{h}_j^*(f)}{S_n^{ij}(f)} \,df 
\end{equation}
 is the usual {inner product} computed over $f_\text{min} = 0.1\,\text{mHz}$ and $f_\text{max} = 10\,\text{mHz}$.
We consider a source to be resolvable if it exceeds the SNR threshold $\rho_{\rm th}=7$~\cite{2024A&A...692A.165T}.

Despite its approximate nature, this iterative subtraction method provides a realistic estimate of the confusion noise,
 comparable to what a global fit pipeline would reconstruct from the data, without requiring a full LISA global fit.
The key approximation is that the algorithm assumes perfect knowledge of the source parameters, and therefore a perfect subtraction. 
Addressing imperfect subtraction and/or source resolvability is deferred to future work.
We do not expect this to significantly impact the results of this work because our inference process takes into account 
uncertainty estimates in the PSD simulation (see Sec.~\ref{sec: GPR} below). 

The PSD is computed with a running mean estimator in order to match likelihood based 
methods~\cite{Karnesis:2021tsh,Santini:2025iuj}. 
In Appendix \ref{sec: mean vs median}, we briefly explore the differences between running mean and running median estimators in this context and 
the relevance of choosing one or the other.

Relative to the original implementation of Ref.~\cite{Karnesis:2021tsh}, we have ported their code to GPUs, which results in large computational speed ups. 
The updated code is made publicly available at Ref.~\cite{Fast_LISA_Subtraction}. \\
Simulating a catalog of $\mathcal{O}(10^7)$ sources requires $\mathcal{O}(1\,\text{min})$
on a single \textsc{NVIDIA A100} 80Gb chip, corresponding to a speed-up of $\sim 10^2$ with respect to the previous CPU implementation.

GB waveforms are generated using the  \textsc{gbgpu} code~\cite{GBGPU_michael_l_katz_2022_6500434,Katz:2022izt}, 
see also Ref.~\cite{Cornish:2007if}. Each waveform is generated in the frequency domain using $N=128$ points around the central frequency $\f$.
We consider a LISA constellation in a rigid equilateral triangle configuration with the nominal $2.5\times 10^6$ km arm 
lengths. In this approximation, the noise covariance matrix is diagonalized by the A, E, and T time-delay-interferometry variables 
\cite{Prince:2002hp,Hartwig:2021mzw,Tinto:2004wu}, where the T channel is insensitive to GWs and therefore is not considered later in the inference.
Off-diagonal correlations between channels are expected under realistic spacecraft orbits~\cite{Baghi:2023qnq,Hartwig:2023pft, 2025JCAP...06..030K}.
Although they are neglected here for simplicity, as is often done in LISA data-analysis studies, our method can be straightforwardly extended to account for such correlations.

\subsection{Coarse-graining of the data}
The output of the subtraction algorithm is a frequency series of the estimated PSD with resolution $\Delta f = 1/\tobs \simeq 10^{-8}\,\mathrm{Hz}$.
This results in $\mathcal{O}(10^{6})$ data points, which is computationally prohibitive for training SBI.
We therefore coarse-grain the data to effectively downsample our computational problem without a significant loss of information~\cite{Caprini:2019pxz,Flauger:2020qyi,Pieroni:2020rob,Georgousi:2022uyt}.
We split the frequency range from $0.1\,{\rm mHz}$ to $10\,{\rm mHz}$ into $N_\text{bins}=100$ 
logarithmically spaced bins and take the average spectrum for each channel
    \begin{align}
      \tilde{S}_n(f_k) &= \frac{1}{N_k}\sum_{i\in \mathcal{I}_k} {S_n}(f_i) \label{eq: coarse graining spectrum}
    \end{align}
    with
    \begin{align}
    f_k &= \frac{1}{N_k} \sum_{i\in \mathcal{I}_k} \,f_i\,, \label{eq: coarse graining frequencies}
    \end{align}
where the index $k = 1, \dots, N_{\rm bins}$ runs over the bins, $\mathcal{I}_k$ denotes the set of frequency indices in the $k$-th bin, and $N_k$ is the number of frequency samples in the $k$-th bin.

We found $N_{\text{bins}}=100$ to be a good compromise for our application, preserving the main spectral features of the PSD while keeping the dimensionality of the data manageable. 
To simplify the notation, in the following we refer to the coarse-grained PSD as $\sconf$, unless stated.

\subsection{Gaussian process regression}\label{sec: GPR}

We further optimize our procedure by interpolating the confusion noise $\sconf$ as a function of the population parameters $\vec{\Lambda}$ using Gaussian process regression (GPR)~\cite{2006gpml.book.....R}.
\review{This choice is primarily motivated by computational limitations: 
although our forward simulator is relatively fast (see Appendix \ref{sec: subtraction algorithm}), generating $\mathcal{O}(10^5)$ simulations required for SBI training would still 
be prohibitive.}

We treat the confusion noise $\sconf$ in the A and E channels as a function of the population parameters $\vec{\Lambda}$ and the frequency $f$ and fit it with two independent GPR models, one for each channel.
This optimization is run over a set of $N=300$ coarse-grained simulations from the subtraction algorithm. More specifically, we denote by $\vec{\rm X}=\{\vec{\rm x}_k\}_{k=1}^{N}$ the training input data, where the vectors
$\vec{\rm x}_k = \{f_k, \vec{\Lambda}_k\}$ are formed by concatenating the coarse-grained frequency array with the population parameters.
We draw the samples $\vec{\Lambda}_k$ from the prior (Table~\ref{tab: priors}) using Latin hypercube sampling~\cite{McKay:1979},
which allows for a more efficient coverage of the parameter space (e.g.~\cite{Taylor:2018iat}).
We further denote by $\vec{\rm y} = \{\avgsconf_k\}_{k=1}^{N}$ the corresponding confusion-noise realizations, which constitute the outputs to be predicted.
One therefore has
\begin{equation}\label{eq: S_n GP}
    S_n(f, \vec{\Lambda}) \sim \mathcal{GP}[\mu(\vec{\rm x}), k(\vec{\rm x}, \vec{\rm x}')\mid \vec{\rm X}, \vec{\rm y}]
\end{equation}
where $\mu(\vec{\rm x})$ defines the mean of the GPR and $k(\vec{\rm x}, \vec{\rm x}')$ the covariance kernel
computed between two different inputs. We adopt
the Matern 1/2 kernel, defined as
\begin{equation}\label{eq: Matern kernel}
    k(\vec{\rm x}, \vec{\rm x}') = \sigma^2 \exp \left[-\frac{d(\vec{\rm x} - \vec{\rm x'})}{\ell} \right]
\end{equation}
\noindent where $d$ indicates the Euclidean distance, and $\ell$ 
and $\sigma$ are free parameters. We empirically found this choice provides better results than other commonly used kernels, including radial basis functions. %

The GPR hyperparameters are optimized by maximizing the marginal log-likelihood~\cite{2006gpml.book.....R}
\begin{equation}\label{eq: GPR log-likelihood}
    \log p(\vec{\rm y} \mid \vec{\rm X}, \ell, \mu, \sigma) \propto -\frac{1}{2} \vec{\rm y}^T K^{-1} \vec{\rm y} - \frac{1}{2} \log \det K \,,
\end{equation}
where $K$ is the kernel covariance matrix with components $K_{ij} = k(\vec{\rm x}_i, \vec{\rm x}_j)$.
Training is regularized by normalizing $\vec{\rm X}$ to zero mean and unit variance, and by placing weakly informative Gamma priors on the kernel parameters:  $\ell, \sigma \sim \Gamma(2,3)$.
Because the inputs are standardized, these priors mildly favor order-unity values while retaining broad support,
thereby regularizing the optimization and preventing overfitting (e.g.\ $\ell \to 0$) or oversmoothing (e.g.\ $\ell \to \infty$).
Out of the $300$ simulations, 10\% are left out for validation.

We use the the \textsc{GPFlow} library~\cite{GPflow2017, GPflow2020multioutput} that  also leverages
GPU acceleration,
and fit two independent models for the A, E channels. Each of these fits
required $\mathcal{O}(2\text{hr})$ of training on a single \textsc{NVIDIA A100} 80Gb GPU at full memory load.

Once each GPR is trained, generating a realization $S_n^{\rm A,E}(f, \vec{\Lambda}')$ requires sampling from the Gaussian prediction provided by the interpolator.
However, if we naively follow Eq.~\eqref{eq: S_n GP}, each frequency bin would be drawn independently from a Gaussian distribution with a large variance, effectively introducing spurious features in the PSD.  
We instead take
\begin{equation}\label{eq: GPR prediction}
  S_n^{\rm A,E}(f, \vec{\Lambda}') = \mu_{\rm GPR}^{\rm A,E}(f, \vec{\Lambda}') + \lambda \, \sigma_{\rm GPR}^{\rm A,E}(f, \vec{\Lambda}'),
\end{equation}
where $\mu_{\rm GPR}^{\rm A,E}(f, \vec{\Lambda}')$ and $\sigma_{\rm GPR}^{\rm A,E}(f, \vec{\Lambda}')$ are the mean and standard deviation of the GPR prediction, respectively,
and $\lambda \in \mathbb{R}$ is drawn from a standard normal distribution, $\mathcal{N}(0,1)$.
Crucially, $\lambda$ is sampled only once and it is kept constant across frequencies. This procedure preserves the overall shape and smoothness of the predicted PSD, with fluctuations induced by $\sigma_\text{\tiny GPR}^\text{A,E}(f, \vec{\Lambda}')$ that are correlated across frequencies.
Incorporating the GPR uncertainty in this way effectively exposes the inference to multiple plausible realizations of the confusion noise associated with the same set of population parameters.
This strategy enhances the robustness of the SBI framework against fluctuations in the estimated PSD.
We note, however, that this approach does not allow for non-smooth features in the PSD, which are expected in realistic settings.
For completeness, we note that $\sigma_{\text{\tiny GPR}}$ also captures uncertainty arising from the finite size of the training set, which turns out to be the dominant source of uncertainty in our GPR model, as 
we will discuss more quantitatively in Sec.~\ref{sec: inference and testing procedure}.

\begin{figure*}
  \centering 
  \hspace*{-0.06\textwidth}
  \includegraphics[width=0.9\textwidth]{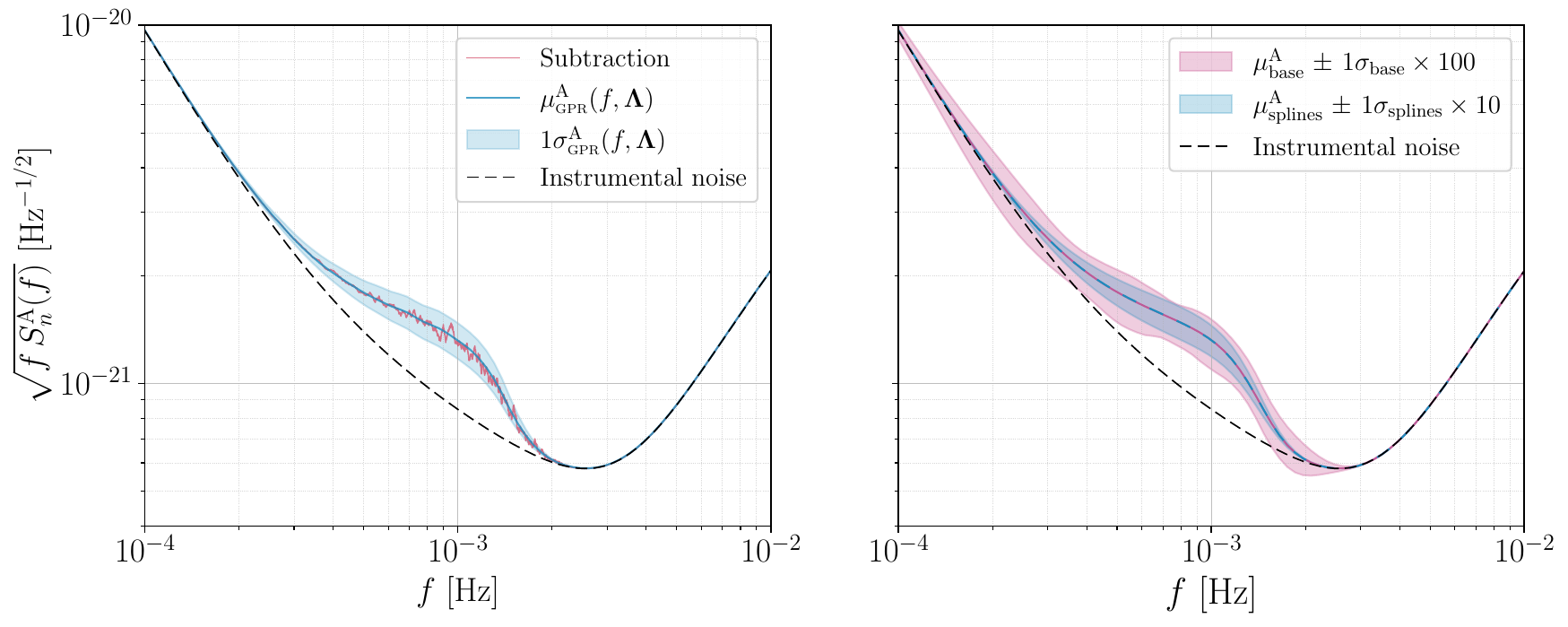}

  \caption{{\it Left.} Comparison between the signal $S_n^\text{A}(f)$ obtained from the subtraction algorithm (red) and the corresponding GPR prediction (blue) 
  for $\vec{\Lambda} = \{\alphaPL = -3.82, \alphaGamma = 2.5, \muGamma = -20.5, \betaGamma = 2, \mu = -21.8, \sigma = 0.24, \corr = 0.85, \Nbinaries = 15\times10^6
\}$.
{\it Right.} MCMC reconstructions of the galactic foreground from the same test simulation 
  using the base model (red) and the splines (blue). The solid lines are given by averaging over posterior draws, while the 
  uncertainity (shaded area) is computed as the $1\sigma$ over the same draws times a factor 100 (10) for the base (spline) model 
  for visualization purposes.
  }
  \label{fig: gpr vs simulation}
\end{figure*}

The left panel of Fig.~\ref{fig: gpr vs simulation} shows a direct comparison between a PSD obtained from the subtraction algorithm and the corresponding GPR prediction 
for $\vec{\Lambda} = \{\alphaPL = -3.82, \alphaGamma = 2.5, \muGamma = -20.5, \betaGamma = 2, \mu = -21.8, \sigma = 0.24, \corr = 0.85, \Nbinaries = 15\times 10^6
\}$. 
This choice is made for testing purposes, in order to produce a GB population qualitatively different from that of Ref.~\cite{Lamberts:2019nyk}.
The mean $\mu_{\text{\tiny GPR}}$ closely tracks the simulation (orange line), while the uncertainty $\sigma_{\text{\tiny GPR}}$ 
fully encompasses the simulation’s variance.

\subsection{Normalizing flow}\label{sec: normalizing flow}
The core of our inference method makes use of normalizing flows~\cite{2019arXiv191202762P}, which we use to
define a surrogate posterior distribution $q_\phi(\vec{\Lambda} |\boldsconf)$ that approximates the true posterior 
$p(\vec{\Lambda} |\boldsconf)$, where we take the data $\boldsconf = \{S_n^\text{A}, S_n^\text{E}\}$
to be a stack of the coarse-grained PSD in the A and E channels.
The surrogate $q_\phi(\vec{\Lambda} |\boldsconf)$ is constructed by means of 
an invertible map $f_\phi$ that transforms it into a standard Normal distribution: 
\begin{equation}\label{eq: flow posterior}
  \begin{split}
  p(\vec{\Lambda}\mid \boldsconf) &\simeq  q_\phi(\vec{\Lambda}\mid \boldsconf) = \\
  &= \mathcal{N}
  \left[f_\phi(\vec{\Lambda}, \boldsconf)\right]
  \left| \det \left(\frac{\partial f_\phi(\vec{\Lambda},\boldsconf) }{\partial \vec{\Lambda}} \right) \right|\,.
  \end{split}
\end{equation}
The map $f_\phi$ is modeled as a concatenation of 16 affine coupling layers~\cite{2016arXiv160508803D} whose 
parameters $\phi$ are optimized 
during the training phase, as implemented in the \textsc{hyperion} sampler~\cite{DeSanti:2024oap, HYPERION}. 
We condition the flow on the data $\boldsconf$ using an embedding network that combines both channels extracting a lower-dimensional representation 
of their relevant features, which is then fed as auxiliary input to each coupling layer; see, e.g., Refs.~\cite{DeSanti:2024oap, Dax:2021tsq}. 
Our embedding module is a convolutional neural network with the architecture summarized in Table \ref{tab: embedding network architecture}.

We jointly train the flow and the neural network by minimizing the Kullback-Leibler (KL) divergence between 
 the target posterior 
 and the flow surrogate, written 
as the negative log-likelihood of the flow:
\begin{align}\label{eq: KL train loss}
\text{KL}( p, q_\phi)
&=
    \mathbb{E}_{[p(\boldsconf)]} \int d\vec{\Lambda}\, p(\vec{\Lambda}| \boldsconf) \log \left[\frac{p(\vec{\Lambda}| \boldsconf) }{q_\phi(\vec{\Lambda}| \boldsconf) } \right]
 \notag \\
  &
\simeq -\frac{1}{N}\sum_{i=1}^{N}\log q_\phi (\vec{\Lambda}^{(i)}| \boldsconf^{(i)}) \;.
\end{align}
We simulate a training dataset composed of $N=10^5$ samples, $\{(\vec{\Lambda}^{(i)}, \boldsconf^{(i)})\}_{i=1}^{N}$, 
where the $\vec{\Lambda}$'s are drawn from the prior $p(\vec{\Lambda})$, and the $\boldsconf$'s are obtained from the GPR predictions.
During training, we augment the dataset by incorporating the GPR uncertainty as described in Sec.~\ref{sec: GPR}, so that the network
is exposed to multiple realizations of the confusion noise for the same population parameters.
Using the same procedure, we generate an additional $10^4$ samples for validation.

\begin{figure}
  \centering
  \vspace{0.14cm}
  \includegraphics[width=\columnwidth]{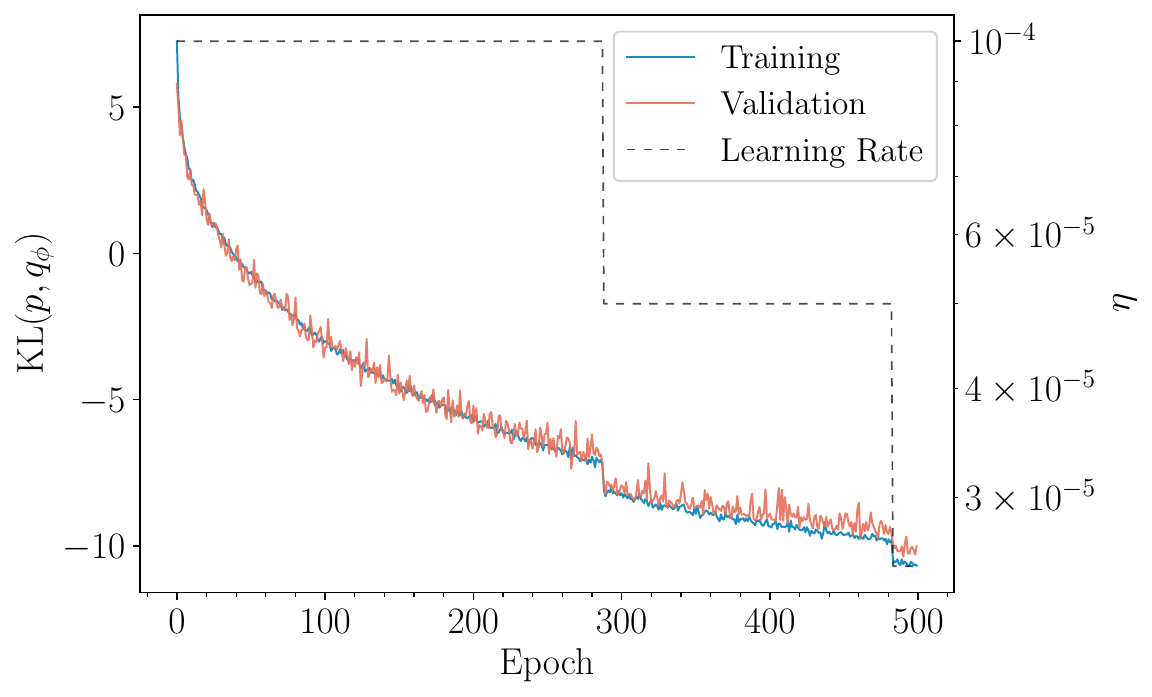}
  \caption{Training (blue) and validation (red) losses as a function of the epoch number (left $y$-axis). 
  The dashed line indicates the learning rate $\eta$ (right $y$-axis).}
  \label{fig: training curve}
\end{figure}

The model is trained for 500 epochs using the Adam optimizer~\cite{Adam_optimizer}, a batch size of 256, and an initial learning rate $\eta = 10^{-4}$, which is reduced by a factor of 2 whenever the validation loss fails to improve for 20 consecutive epochs. 
The number of epochs is chosen empirically to avoid overfitting, which is monitored by tracking the training and validation losses, as shown in Fig.~\ref{fig: training curve}.
No significant overfitting is indeed observed, 
as the validation loss closely follows the training loss, indicating good convergence of the optimization process. 
The sharp drops in both losses correspond to reductions in the learning rate.

\begin{table}[!t]
  \renewcommand{\arraystretch}{1.3}  \centering
    \begin{tabular}{c@{ \hspace{5pt} }c @{ \hspace{5pt} } c}
    \hline
    \text{Layer} & \text{Parameters} & \text{Output Shape} \\
    \hline
    \hline
    Input & -- & $(B, 2, 100)$ \\
    BatchNorm1d & & $(B, 2, 100)$ \\
    Conv1d & kernel\_size = 10& $(B, 32, 91)$ \\
    ELU & -- &  \\
    BatchNorm1d & & $(B, 32, 91)$ \\
    Conv1d & kernel\_size = 10 & $(B, 64, 82)$ \\
    ELU & -- & \\
    BatchNorm1d & & $(B, 64, 82)$ \\
    Conv1d & kernel\_size = 10& $(B, 128, 73)$ \\
    ELU & -- & \\
    BatchNorm1d &  & $(B, 128, 73)$ \\
    Flatten & -- & $(B, 9344)$ \\
    Linear & & $(B, 128)$ \\
    ELU & -- &  \\
    Linear & & $(B, 128)$ \\
    ELU & -- & \\
    \hline
    \end{tabular}
    \caption{Architecture of the embedding network. The input shape corresponds to the A and E channels of the coarse-grained 
    spectrum; $B$ indicates the batch size.}
    \label{tab: embedding network architecture}
  \renewcommand{\arraystretch}{1}
  \end{table}

\subsection{Internal validation}\label{sec: inference and testing procedure}
We internally validate our end-to-end pipeline to demonstrate that it achieves accurate Bayesian coverage.
To this end, we generate posterior samples from the surrogate posterior $q_\phi(\vec{\Lambda} \mid \boldsconf(f))$ 
by drawing samples from a standard normal distribution and applying the inverse of the map $f_\phi$ defined in Eq.~\eqref{eq: flow posterior}.
Since each $\boldsconf(f)$ is generated via the GPR model, 
we account for its uncertainty by combining posterior samples obtained from multiple realizations, in analogy with the data augmentation procedure.
We indicate by $\boldsconf^{(\lambda)}(f)$ the PSD generated using the same $\mugpr$ and $\sigmagpr$ but different $\lambda$ [cf.~Eq.~\eqref{eq: GPR prediction}].  One has
\begin{equation}\label{eq: posterior samples combination}
  \begin{split}
  p(\vec{\Lambda} \mid \{\boldsconf(f)\}) &= \int d\lambda\,p(\vec{\Lambda} \mid \boldsconf^{(\lambda)}(f), \lambda)\,p(\lambda)\simeq \\
  &\simeq \frac{1}{N_\lambda}\sum_{n=1}^{N_\lambda} q_\phi(\vec{\Lambda} \mid \boldsconf^{(\lambda)}(f))
  \end{split}
  \end{equation}
where $N_\lambda$ denotes the number of different realizations. \\
We treat each foreground realization as an independent input and
average the resulting posteriors, i.e. we approximate the integral in Eq.~\eqref{eq: posterior samples combination} by stacking $10^3$ posterior 
samples from $N_\lambda=10^3$ PSD draws. 
In practice, this procedure marginalizes over the GPR prediction, 
thereby improving the robustness of our inference. 
We emphasize that this is exactly the approach we would adopt when combining posterior samples from a global fit analysis,
where one would have a set of curves drawn from the posterior for the parameters of $\sgal$ and $\sinstr$.

\begin{figure}
  \centering
  \includegraphics[width=\columnwidth]{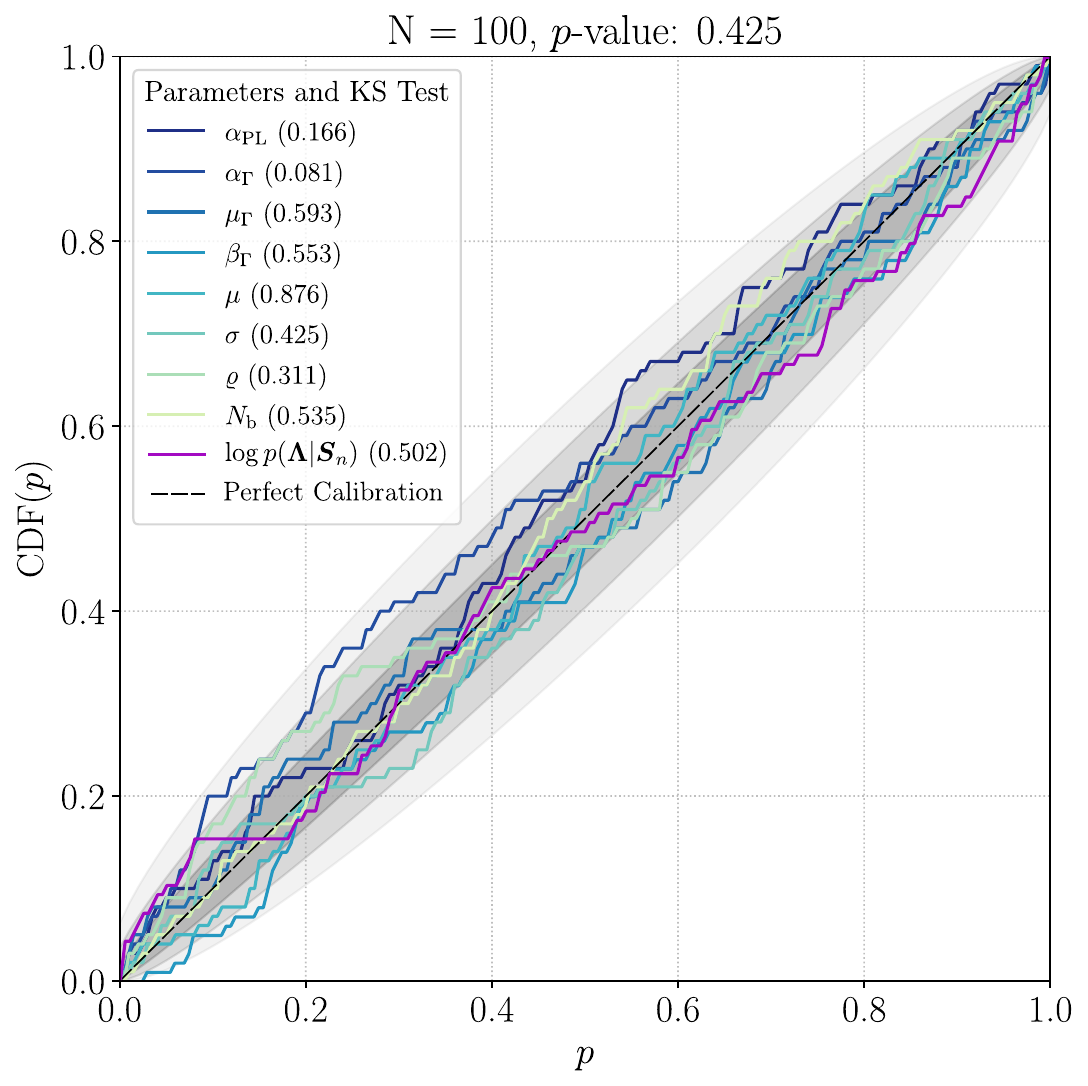}
  \caption{Probability-probability plot obtained from 100 independent simulations in the test set.
The shaded regions correspond to the $1\sigma$, $2\sigma$, and $3\sigma$ confidence intervals for a binomial distribution.
The legend reports the KS-test $p$-value for each \review{$\vec{\Lambda}$} parameter \review{and the logarithm of the posterior density $\log p(\vec{\Lambda} | \boldsconf)$. The combined $p$-value of 0.425
is computed with Fisher's method using only the marginals.}}
  \label{fig: PP plot}
\end{figure}

We first validate the statistical calibration of our inference framework by analyzing 100 independent test simulations, 
with $\vec{\Lambda}$ drawn from the prior.
The results are summarized in the probability–probability (\pp) plot shown in Fig.~\ref{fig: PP plot}, 
where we show the cumulative distribution functions (CDFs) of the fraction of true values contained within a given credible interval for each parameter.
All \pp curves are consistent with the diagonal within statistical fluctuations, where a diagonal \pp plot indicates perfect Bayesian coverage.
\review{To further assess statistical coverage, we compute the \pp curve for the posterior density $p(\vec{\Lambda} |\boldsconf)$.
This approach allows us to probe the calibration of the \emph{full} n-dimensional posterior, 
rather than restricting the assessment to its marginal projections.
The density is obtained directly from the flow $q_\phi(\vec{\Lambda} | \boldsconf)$.}
Figure~\ref{fig: PP plot} also reports the $p$-values resulting from Kolmogorov–Smirnov (KS) tests.
Using Fisher’s method, \review{we obtain a combined $p$-value of $0.425$ for the marginals, which---together with the $p$-value of $0.502$ for $\log p(\vec{\Lambda} | \boldsconf)$---suggests}
 no significant evidence to reject the null hypothesis of perfect Bayesian coverage, indicating that our inference framework is well calibrated.

\subsection{External validation}\label{sec: inference and testing procedure 2}

We further test our model under more realistic conditions by performing inference on MCMC reconstructions of the Galactic foreground 
from a single realization of the LISA data stream $d(f)$, which contains both instrumental noise and the sum of unresolved GBs 
obtained via the subtraction algorithm.
We fit parametric models for $\sgal$ as in Ref.~\cite{Santini:2025iuj}, while keeping the instrumental noise fixed, adopting two approaches of increasing complexity.

We first assume the model from Ref.~\cite{Karnesis:2021tsh,Robson:2018ifk},
\begin{equation}\label{eq: confusion noise}
    \sgal = \frac{A}{2} f^\gamma e^{-(f/f_1)^\alpha}
    \left[1 + \tanh \left(\frac{f_{\rm knee}-f}{f_2}\right) \right]
\end{equation}
with parameters $\vec{\xi} = \{ A, \, \gamma, \, f_1, \, \alpha, \, f_{\rm knee}, \, f_2\}$, which we refer to as our ``{base model},'' 
and run the inference.
Here, $\gamma$ is the spectral index (with $\gamma = -7/3$ for a population of quasi-circular GBs inspiralling solely due to 
GW emission~\cite{2001astro.ph..8028P}), $f_1$, $f_2$, and $\alpha$ are scaling parameters related to the population, 
source-resolvability thresholds, and observational time, and $f_{\rm knee}$ is the high-frequency cutoff, 
reflecting the fact that sources at higher frequencies are louder and therefore more likely to be resolved.

Next, we fix the parameters of the {base model} to their maximum likelihood values $\bar{\vec{\xi}}$ 
and add a flexible spline component~\cite{Santini:2025iuj}. 
The Galactic foreground is therefore modeled as
\begin{equation} \label{eq: spline confusion noise}
    S_n(f, \, \vec{\xi}_{\rm spline}) = S_{n, \rm base}(f, \, \bar{\vec{\xi}}) \times 10^{\mathcal{S}(f, \, \vec{\xi}_{\rm spline})} \,,
\end{equation}
where $\vec{\xi}_{\rm spline}$ represents the number, positions, and amplitudes of the spline knots, and 
$\mathcal{S}(f, \, \vec{\xi}_{\rm spline})$ is a linear spline interpolation function. 
We refer to this modeling approach as the ``{spline model}.'' 
The spline component is a linear interpolant in which the number of knots is not fixed a priori, 
but determined by the data using a reversible-jump MCMC algorithm~\cite{Green:1995mxx}. 
We place uniform priors on the spline amplitudes between $-1$ and $1$, allowing for variations of up to one order of magnitude around the base model.
We use the  \textsc{Eryn} sampler~\cite{Karnesis:2023ras}, and the Whittle likelihood defined as~\cite{whittlelikelihood}:
\begin{equation}\label{eq: whittle likelihood}
  \begin{split}
    &\log \mathcal{L}(d \mid \vec{\xi}) = \\
    &=- \frac{1}{2} \sum_{j\in\{A,E\}} \sum_{i=1}^{N_f} \left[ \frac{|d_{ji}|^2}{S_{n,i}^j(\vec{\xi})} + \log S_{n,i}^j(\vec{\xi}) + \log 2\pi \right]\,.
  \end{split}
\end{equation}
We refer the reader to Ref.~\cite{Santini:2025iuj} for details on this spline-based modeling approach and the adopted reversible-jump MCMC algorithm.
For external validation purposes, we then feed the curves $\boldsconf(f)$, drawn from the MCMC posterior distributions of both reconstructions, into our normalizing flow.

An example of such a reconstruction is shown in the right panel of Fig.~\ref{fig: gpr vs simulation}, 
where we reconstruct the PSD for a test simulation with 
$\vec{\Lambda} = \{\alphaPL = -3.82, \alphaGamma = 2.5, \muGamma = -20.5, \betaGamma = 2, \mu = -21.8, \sigma = 0.24, \corr = 0.85, \Nbinaries = 15\times10^6 \}$.
Both the base and spline models provide very accurate reconstructions of the injected signal, 
with the base model showing an uncertainty about an order of magnitude lower than the spline model, which is expected given its larger flexibility. 
Comparing against the GPR prediction shown in the left panel of Fig.~\ref{fig: gpr vs simulation} illustrates that both reconstructions capture the smooth features of the PSD.
However, the GPR uncertainty is about an order of magnitude larger than that of the spline model reconstruction.
This implies that, in the network predictions, the systematic uncertainty introduced by the GPR at the training stage
dominates over the statistical uncertainty. This is likely due to the limited size of the training simulations used for the GPR, 
and could, in principle, be mitigated by increasing their number at the expense of more computational resources and longer processing times.
We postpone such an exploration to future work.

\section{Results }\label{sec: results}
\subsection{Inference on simulations}

\begin{figure*}[p]
  \centering
  \includegraphics[width=\textwidth]{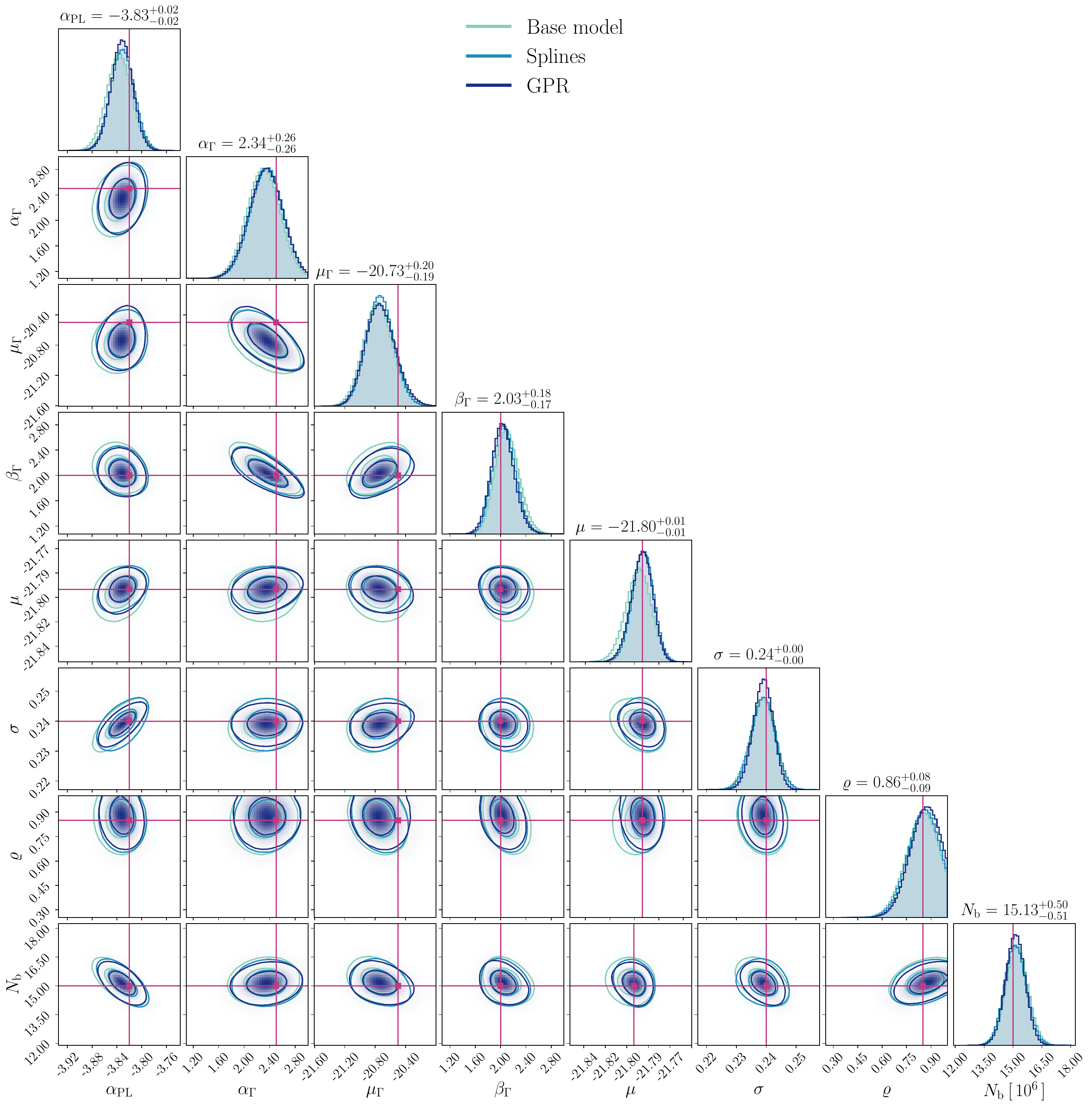}
  \caption{Inference results for the simulated catalog of
  Fig.~\ref{fig: gpr vs simulation}. We compare posteriors inferred in three different scenarios: direct conditioning on the GPR simulation (blue), 
  spline model reconstruction of the PSD (light blue), and base model reconstruction of the PSD  (green). Red lines lines indicate the injected values. 
  Contours refer to the 50 -- 90\% credible intervals; 
  estimators reported above the 1D marginals refer to medians and 16$^{\rm th}$-- 84$^{\rm th}$ percentiles of the GPR-based inference. }
  \label{fig: gpr corner}
\end{figure*}

\begin{figure}
  \centering
  \includegraphics[width=\columnwidth]{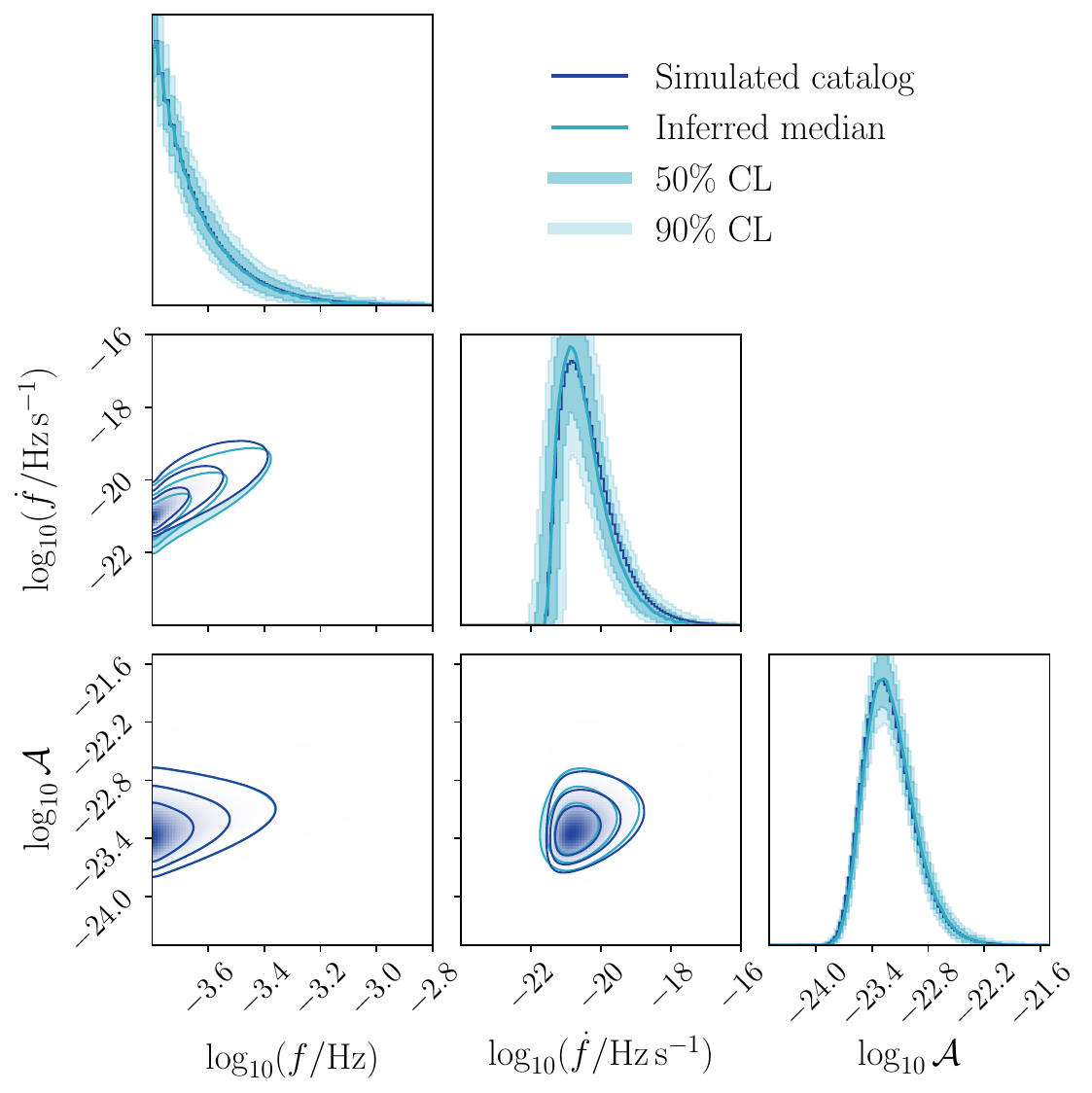}
  \caption{Source catalog reconstruction from the (GPR) posterior samples of Fig.~\ref{fig: gpr corner}.
The injected catalog is shown in blue, while the posterior predictive median and its 50--90\% credible intervals are shown in cyan. 
Contours correspond to the 39--67.5--86\% credible regions. 
 }
  \label{fig: reconstructed catalog}
\end{figure}

\begin{figure*}
  \centering
  \includegraphics[width=\textwidth]{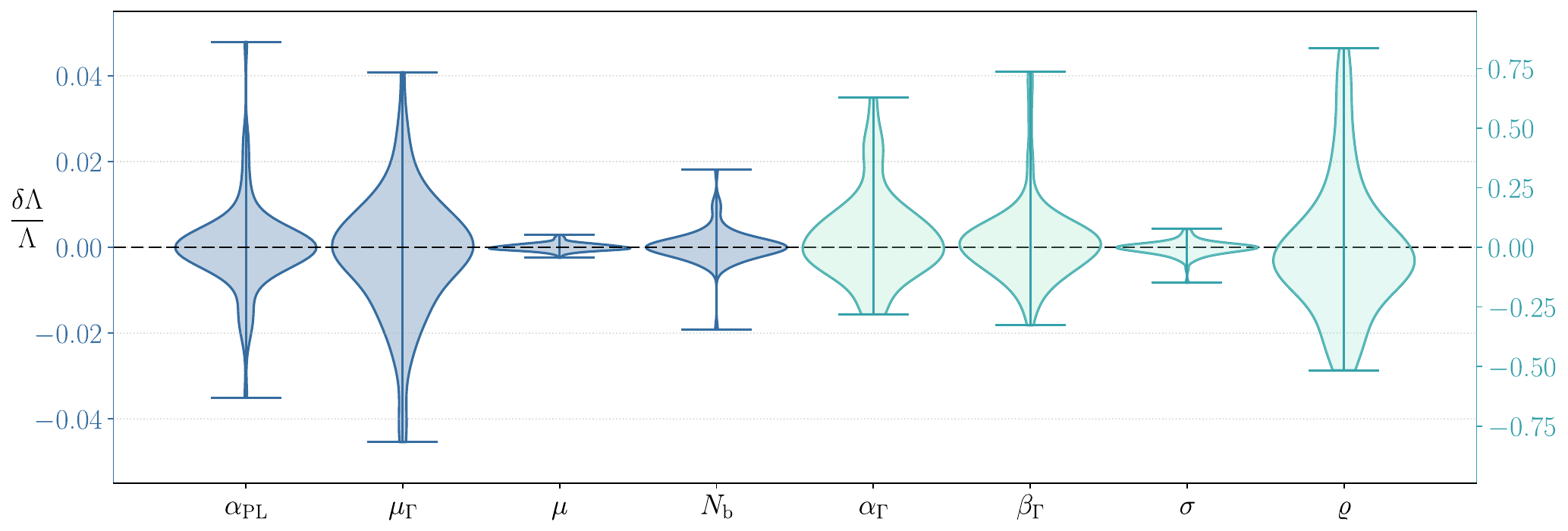}
  \caption{Distribution of the relative error,
  $\delta \Lambda / \Lambda$, on the median of the 1D posterior marginals, for each parameter, across 100 independent
  test simulations. 
  Left axis (blue) refers to $\alphaPL$, $\muGamma$, $\mu$, and $\Nbinaries$, while right axis (green) refers to $\alphaGamma$, $\betaGamma$, $\sigma$, and $\corr$.
  }
  \label{fig: relative error violin}
\end{figure*}

Figure~\ref{fig: gpr corner} shows our inference results for the same simulation already presented in Fig.~\ref{fig: gpr vs simulation}. 
We perform inference using both the internal and external validation methods outlined previously, 
i.e., conditioning on both the GPR draws and the MCMC reconstructions.
We combine posterior samples as in Eq.~\eqref{eq: posterior samples combination}, averaging over 
$N_\lambda = 10^3$ independent draws of $\boldsconf(f)$ obtained either from the GPR model or from the MCMC fit. 
For the GPR case, these curves are sampled directly from the predictive distribution [cf. Eq.~\eqref{eq: GPR prediction}],
whereas for the base and spline reconstructions we first summarize the MCMC posterior by its mean and standard deviation at each frequency.

The two approaches are in excellent agreement: the posterior credible regions largely overlap, 
and the 1D marginals have comparable widths. The only appreciable differences are a slight broadening 
in $\alphaPL$ and $\mu$ for the base-model reconstruction. This consistency reflects the fact that all 
three reconstructions capture the same smooth PSD features (Fig.~\ref{fig: gpr vs simulation}), 
thereby validating our external inference procedure, including the approach we use to marginalize over the GPR uncertainty.

All parameters are well recovered, with nearly Gaussian marginals centered on the true values. Correlations are generally weak, 
with notable anti-correlations in $(\alphaPL, \Nbinaries)$, $(\alphaGamma, \betaGamma)$, and $(\alphaGamma, \muGamma)$, and positive correlations in 
$(\corr, \Nbinaries)$ and $(\alphaPL, \sigma)$.
In particular, the negative correlation between $\alphaPL$ and $\Nbinaries$ is consistent with the behavior described in Sec.~\ref{sec: GPR}. 
As discussed for Fig.~\ref{fig: gpr predictions}, the high-frequency amplitude of $\sgal$ increases for flatter power laws (larger $\alphaPL$) 
or for a larger number of sources. We also find mild correlations involving $\corr$, the Gaussian-copula parameter controlling the joint 
$p(f,\fdot)$ distribution. This suggests that the information carried by the $f$--$\fdot$ coupling, encoded in the overall foreground 
shape, is weakly degenerate with the specific marginal distributions. Extending the analysis to include additional effects 
[Eq.~\eqref{eq: fdot astrophysical}] is left for future work.

Figure~\ref{fig: reconstructed catalog} shows the posterior-predictive reconstruction of the source catalog based on the posterior samples 
of Fig.~\ref{fig: gpr corner} for the inference on the GPR data. We draw $10^3$ catalogs from our $\ppop(\theta|\vec{\Lambda})$ 
and report the median and
50--90\% credible intervals alongside the injected catalog. The agreement is excellent, with the injected
catalog lying within $1\sigma$ of the reconstruction.

Complementing the statistical coverage test in Fig.~\ref{fig: PP plot}, we summarize in Fig.~\ref{fig: relative error violin} 
the accuracy of our inference by computing the relative error
on the posterior medians across our test simulations. For
$\Lambda = \{\alpha_\text{PL}, \mu_\Gamma, \mu, \Nbinaries \}$ we find $\delta \Lambda / \Lambda_\text{true} \lesssim 4\%$. 
The small relative error achieved by $\Nbinaries$ indicates that the
foreground alone carries enough information to tightly constrain the total number of Galactic binaries (resolved and unresolved),
thereby informing the Galactic structure and star-formation history~\cite{Korol:2021pun, Georgousi:2022uyt}. 
The parameter $\corr$ is measured within 50\%, suggesting that robustly constraining the correlation in $f$--$\fdot$ and therefore placing constraints 
on astrophysical effects in GBs evolution \cite{2024A&A...692A.165T,Delfavero:2024zyl,Georgousi:2022uyt, 2026arXiv260211765M, 2002MNRAS.329..897H} 
may require incorporating information from resolved sources.

\subsection{Inference on astrophysical catalog}\label{sec: astro catalog inference}
\begin{figure*}[p]
  \centering
  \includegraphics[width=\textwidth]{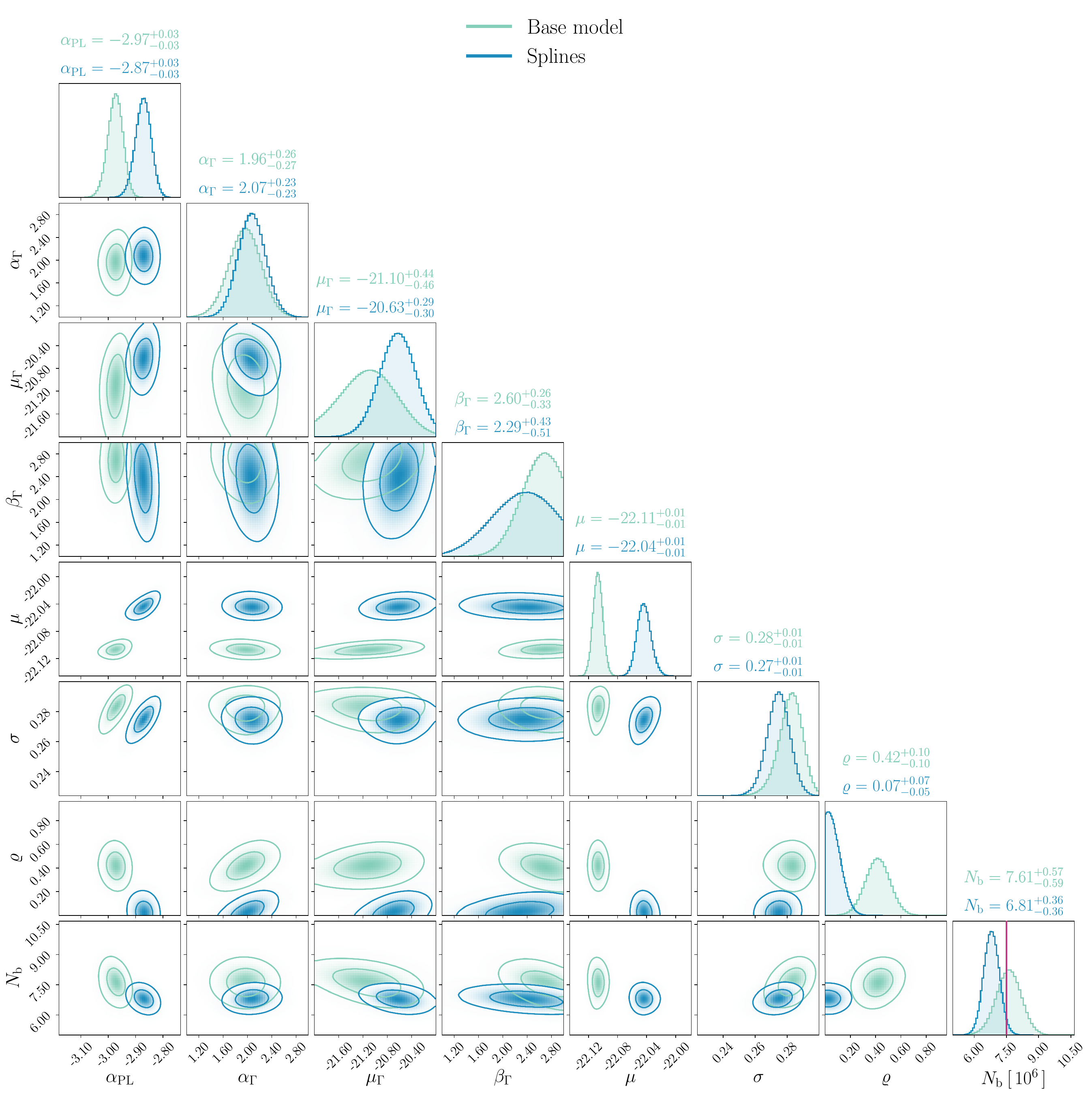}
  \caption{Inference on the astrophysical catalog of Ref.~\cite{Lamberts:2019nyk} using the base-model (green) and spline (light blue) reconstructions of the foreground.
  The vertical red line indicates the total number of sources in the catalog.
  Contours refer to the 50 -- 90\% credible intervals; 
  estimators reported above the 1D marginals refer to medians and 16$^{\rm th}$-- 84$^{\rm th}$ percentiles.
  }
  \label{fig: astro cat inference corner}
\end{figure*}

\begin{figure*}
  \centering
  \includegraphics[width=0.95\textwidth]{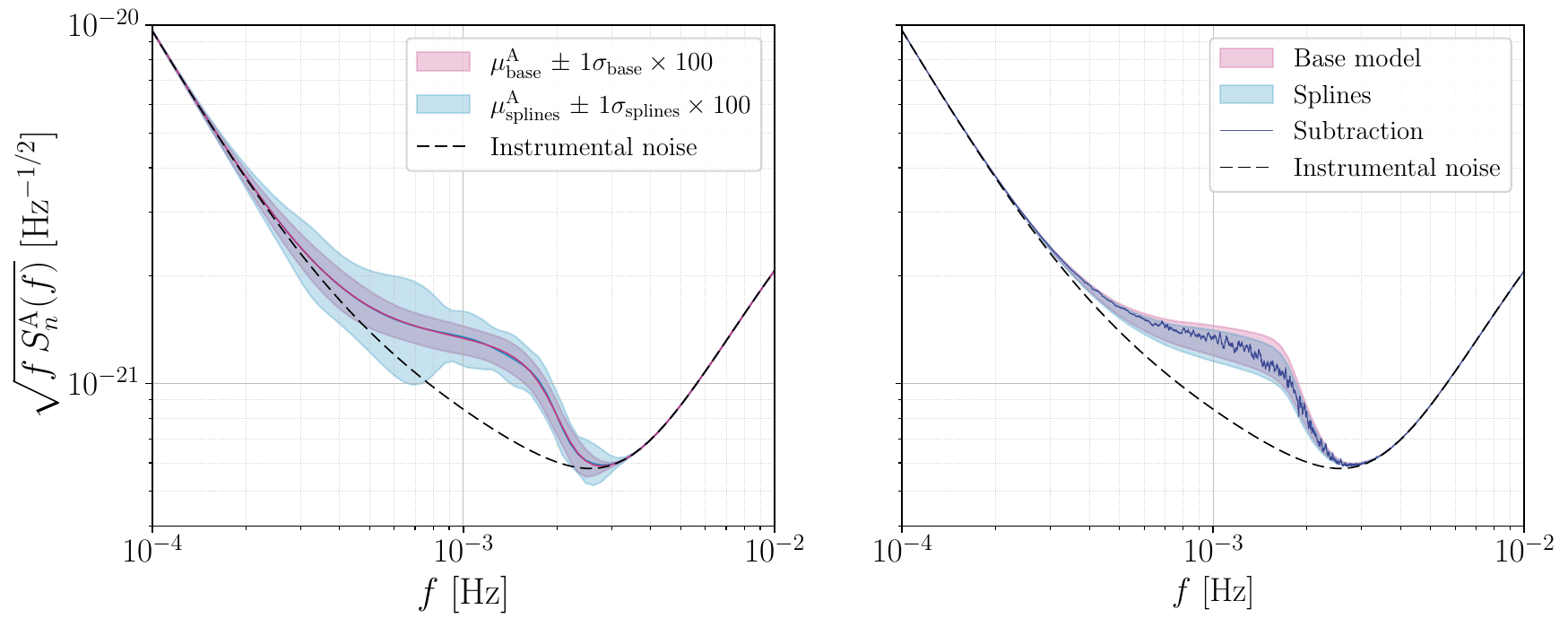}
  \caption{{\it Left.} MCMC reconstructions of the galactic foreground from the astrophysical catalog of Ref.~\cite{Lamberts:2019nyk} 
  using the base model (red) and the splines (blue). Note that confidence intervals have been exaggerated by a factor of 100 for readability purposes.
  {\it Right.} Comparison between the true $\avgsconf$ obtained from the subtraction algorithm (solid line)
  and the reconstructed one from the posterior samples of Fig.~\ref{fig: astro cat inference corner} using splines (blue) and base model (red). 
  The shaded regions are obtained by drawing $10^2$ samples from the posterior and generating the corresponding foreground from the GPR
  [cf. Eq.~\eqref{eq: GPR prediction}].
  } 
  \label{fig: astro sgal reconstruction}
\end{figure*}

We now apply our inference framework to the astrophysical catalog of Ref.~\cite{Lamberts:2019nyk}. 
In this case, the true population hyperparameters are not known a priori, except for the total number of sources,
$\Nbinaries = 7.5 \times 10^6$. We therefore perform inference using only the base-model and spline MCMC 
reconstructions of the foreground, and compare the results in Fig.~\ref{fig: astro cat inference corner}. 
In contrast to the previous case, the inferences are not fully consistent with each other: the posteriors 
only mildly overlap for most parameters, with noticeable discrepancies in $\alphaPL$ and $\mu$. 
This may indicate a systematic difference between the two PSD reconstructions for this catalog, which 
are shown in the left panel of Fig.~\ref{fig: astro sgal reconstruction}.

\begin{figure*}
  \centering 
  \includegraphics[width=\columnwidth]{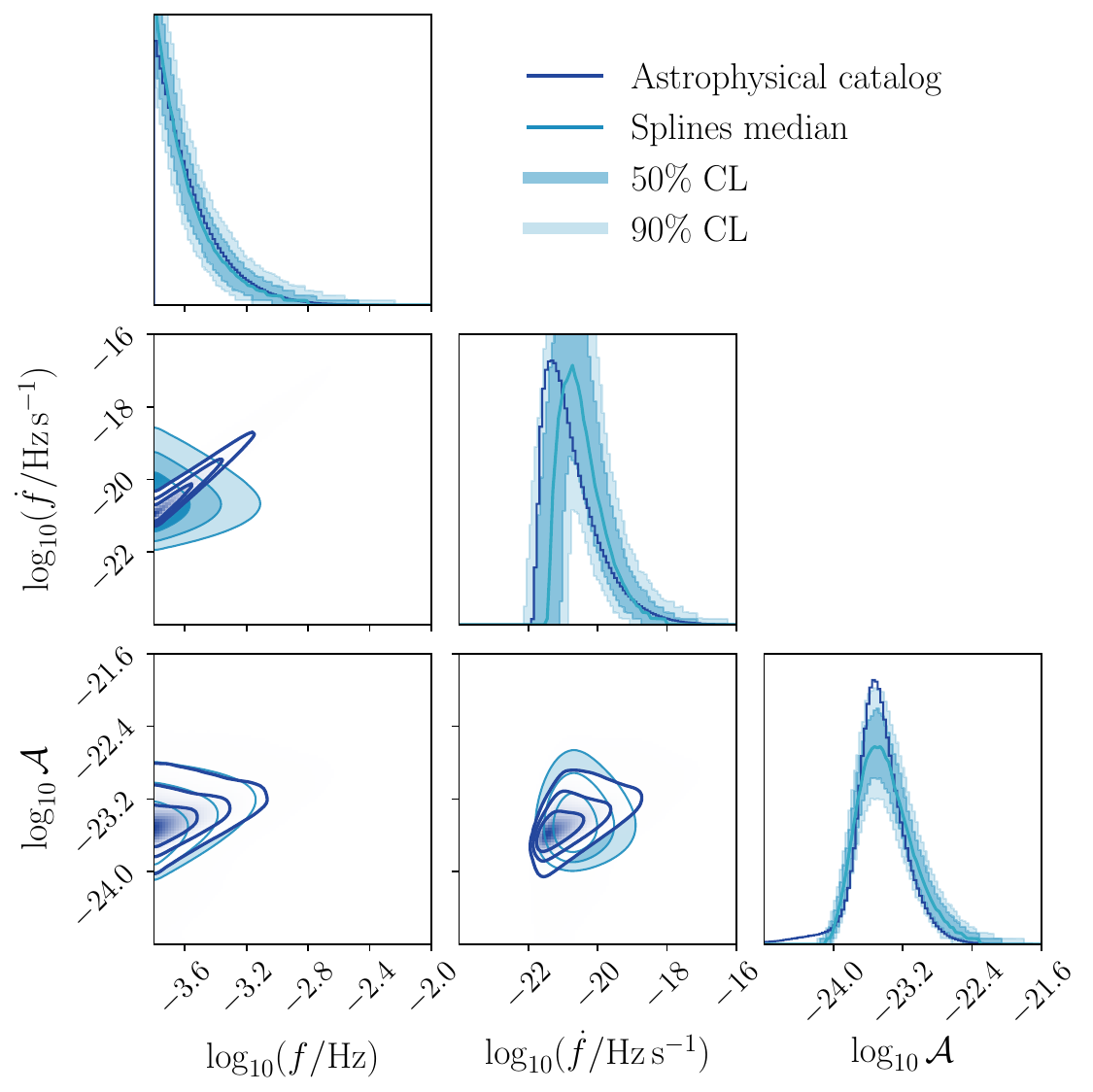}
    \includegraphics[width=\columnwidth]{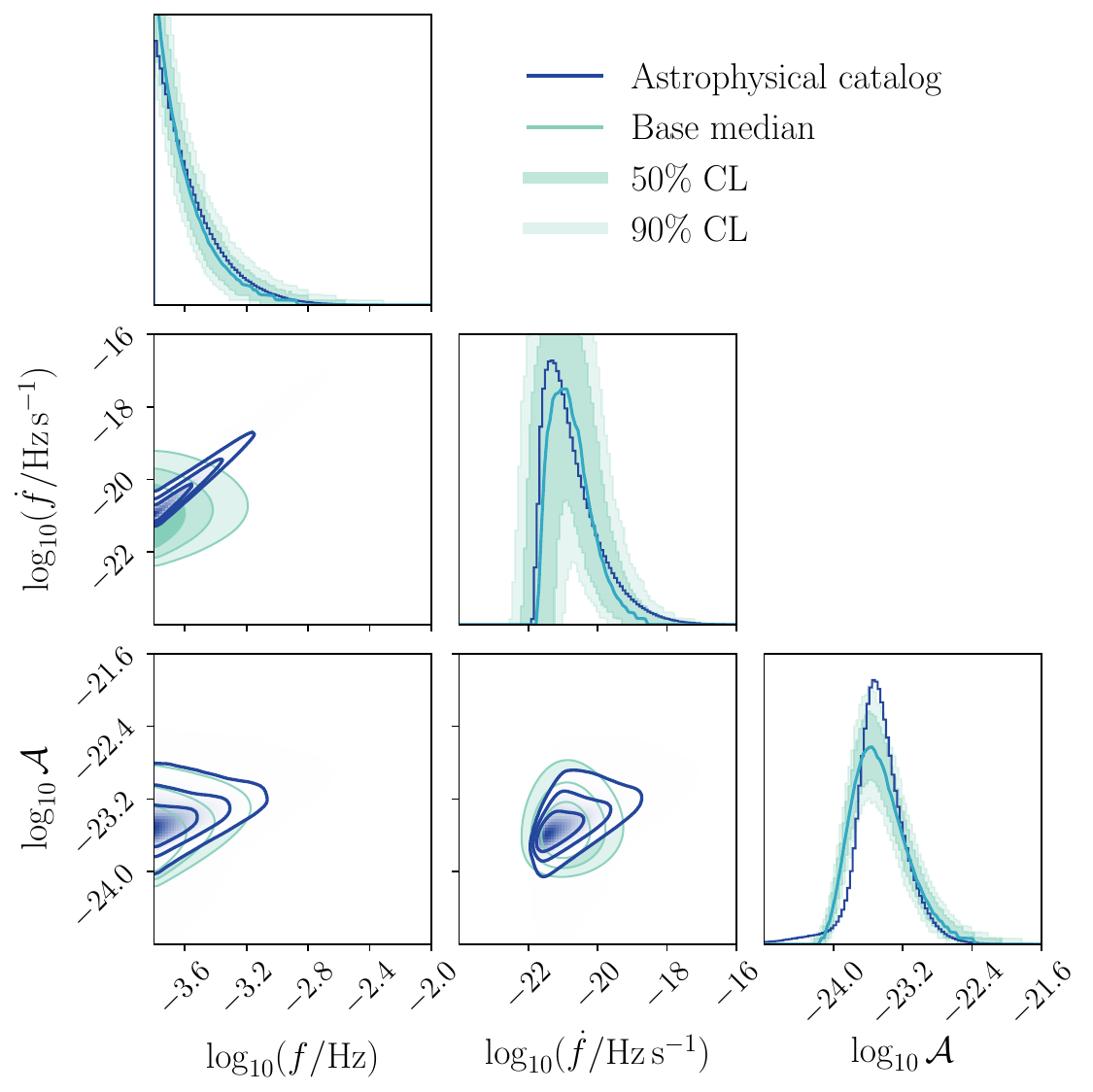}

  \caption{Source catalog reconstruction from the posterior samples of Fig.~\ref{fig: astro cat inference corner}, 
using the spline model (left) and the base model (right). The injected catalog is shown in blue, 
while the posterior predictive median and its 50--90\% credible intervals are shown in cyan (green) 
for the spline (base-model) inference.  Contours refer to the 39--67.5--86\% credible regions.}
  
  \label{fig: astro catalog reconstruction}
\end{figure*}

With the base model we find $\Nbinaries = 7.61^{+0.57}_{-0.59} \times 10^6$, while with the splines we find $\Nbinaries = 6.81^{+0.36}_{-0.36} \times 10^6$.
The base model estimate is consistent with the expected value within uncertainties, whereas the spline-based estimate is biased low. 
At the same time, the spline reconstruction favors a higher $\alphaPL$ due to the anti-correlation 
between these parameters.\\
Other notable differences include the correlation parameter $\corr$, which is favoured to be greater for the base model ($\corr = 0.42^{+0.10}_{-0.10}$)
but almost zero for the spline reconstruction ($\corr = 0.07^{+0.07}_{-0.05}$), and $\mu$, which is shifted to higher values for the spline fit.
The left panel of Fig.~\ref{fig: astro sgal reconstruction} shows that the foreground reconstruction with the splines
exhibits small-scale fluctuations, especially at low frequency, that are absent in the smoother base-model fit. 
In this case, the MCMC reconstructions differ qualitatively from the GPR-based training curves, which explains the observed biases when applying the flow. 

Despite these differences in the posteriors, the posterior-predictive marginals of the catalog remain broadly compatible with the injected
population (Fig.~\ref{fig: astro catalog reconstruction}), even though we observe a bias in the $\fdot$ distribution for the spline fit, 
and alternatively in $\amp$ distribution for the base model. The joint $f$--$\fdot$ distribution 
is instead poorly recovered by both reconstructions, mainly due to an underestimation of $\corr$\review{, which might originate from 
two concurrent effects: the intrinsic difficulty in constraining the correlation parameter, 
and the out-of-distribution features of the MCMC reconstructions.
Indeed, internal validation tests (Fig.~\ref{fig: relative error violin}) indicate that the correlation parameter 
appears to be the least well constrained parameter, with the observed bias consistent with the level of statistical 
uncertainty intrinsic to the trained model. For a well-calibrated model, we expect this intrinsic uncertainty to manifest as 
a broadening of the posterior, rather than a bias. 
Since our model does not show evidence of miscalibration when applied to in-distribution data (Fig.~\ref{fig: PP plot}), we therefore interpret the observed 
bias mainly as a consequence of the out-of-distribution features of the MCMC reconstructions, rather than a poor convergence of the network.}

To further investigate the differences between the two inferences,
we compare the reconstructed foregrounds from the posterior samples of Fig.~\ref{fig: astro cat inference corner}. 
To this end we draw $10^2$ samples from the posteriors and generate the corresponding 
foregrounds using the GPR model [cf. Eq.~\eqref{eq: GPR prediction}] 
and show the resulting 90\% credible regions in the right panel of Fig.~\ref{fig: astro sgal reconstruction}.
We find that both reconstructions are consistent with injection, as the $S_n^\text{A}$ from the subtraction algorithm 
lies within the 90\% credible region.
More in detail, both reconstructions overlap at $f\gtrsim 2$ mHz, i.e. around the knee of the ``bump,'' but differ at lower frequencies, 
where the spline reconstruction is slightly lower than the base model one. 
As in Fig.~\ref{fig: gpr vs simulation}, the GPR uncertainty
remains larger than the MCMC one and confirms to be the dominant source of uncertainty in the network predictions. 
The most direct mitigation strategy remains increasing the number of training simulations; 
we might as well expect that in more realistic analyses the two uncertainities to become more comparable, in particular, when accounting 
for the imperfect knowledge of instrumental noise.

The dicrepancy in the PSD envelope between the spline MCMC reconstruction and the GPR-based 
one can further provide
a qualitative diagnostic of out-of-distribution behavior of the former.
Our current understanding is that the small-scale fluctuations in the PSD, captured by the splines,
around 1mHz (left panel of Fig.~\ref{fig: astro sgal reconstruction}) 
make it out of distribution 
for the network, compared to our simulations (see Sec.~\ref{sec: GPR}).
This is also the region 
most sensitive to changes in $\alpha_{\rm PL}$ and $N_b$ (see, e.g., Fig.~\ref{fig: gpr predictions}).
By contrast, the base model is smooth and leads to a better reconstruction of the astrophysical population, although it may not have sufficient 
flexibility to reproduce the PSD generated by this population. 
At the same time, the parametrization chosen for the amplitude distribution 
could be causing a mismatch in its reconstruction, 
since the population of Ref.~\cite{Lamberts:2019nyk} exhibits a tail toward lower amplitudes due to sources 
in satellites. Future work will investigate the impact of these different aspects and how to mitigate them, 
enhancing the reliability of this framework on global-fit outputs.

\section{Conclusions}\label{sec: conclusions}
We presented a simulation-based inference framework to recover population properties of GBs from the confusion-noise PSD alone.
This targets a key challenge for LISA data analysis: extracting population information accounting for unresolved sources, coupled to resolved ones in
global fit outputs.
Starting from a parametric ansatz for the population model $\ppop(\vec{\theta}|\vec{\Lambda})$ in the observable space, with
$\vec{\theta} = \{f,\fdot,\amp\}$, we use a forward simulator to map population parameters to $\sgal$ and train a neural posterior estimator
to invert that mapping. From a hierarchical inference perspective, our approach is a crucial step towards modeling the likelihood 
$p(d|\{\vec{\theta}_i\}, S_\text{gal} , \vec{\Lambda})$ \cite{2026arXiv260104168T}, over resolved sources,
foreground, and population parameters.

The feasibility of this study crucially relies on the optimized implementation of the subtraction algorithm~\cite{Karnesis:2021tsh}, 
which allows the simulation of stochastic foregrounds from catalogs with $\mathcal{O}(10^7)$ sources within a hundred seconds on a single GPU.
This performance makes large-scale simulation campaigns practical and opens the door to a wide range of applications in LISA data analysis, 
from population inference to studies of detectability and parameter estimation of other stochastic gravitational-wave backgrounds. The optimized implementation of the substraction algorithmn developed for this paper is made publicly available at Ref.~\cite{Fast_LISA_Subtraction}.

On the modeling side, we put forward an agnostic parametrization for $\ppop(\vec{\theta}|\vec{\Lambda})$, informed by the astrophysical catalog of Ref.~\cite{Lamberts:2019nyk}.
We relax the relation between $f$--$\fdot$, by adopting a Gaussian copula description, instead of assuming a GW-driven inspiral as in Eqs.~\eqref{eq: fdot gw}--\eqref{eq: fdot astrophysical}.
This choice minimizes astrophysical assumptions and mitigates potential 
modeling systematics, while retaining sufficient flexibility to capture a variety of astrophysical scenarios. We showed that our parametrization
is able to capture the main features of the astrophysical catalog, while it can be easily and
straightforwardly extended to include additional components, such as astrophysical contributions to $\fdot$ [cf.~Eq.~\eqref{eq: fdot astrophysical}], 
negative-$\fdot$ sources~\cite{2024A&A...692A.165T}, or multiple sub-populations. 
More complex scenarios, including the presence of Milky Way satellites~\cite{Pozzoli:2024wfe, Korol:2020hay}, could also be incorporated
and may leave distinctive imprints on the shape of the confusion foreground~\cite{Georgousi:2022uyt, Benacquista:2005tm}. 
A systematic investigation of these effects is however beyond the scope of this work and is left to future studies.

Using an extensive validation set of simulations, we demonstrated that the confusion noise encodes measurable information about the underlying population. 
In particular, we observed that several population parameters can be recovered with relative uncertainties of a few percent, 
with the total number of sources $\Nbinaries$ emerging as one of the best constrained parameters. We further tested the robustness of the framework 
in a more realistic setting by performing inference on global-fit-like outputs, showing that meaningful population constraints can still be obtained.
We emphasize, however, that the accuracy of the inferred population requires consistency between the PSD reconstructions used at 
inference time and those employed during training, as it turned out to be extremely delicate.
As shown by our tests on astrophysical catalogs, mismatches in the smoothness or spectral features of the reconstructed foreground can introduce biases, 
highlighting the importance of a coherent treatment of PSD modeling throughout the pipeline.\\
\noindent In particular, simulating the spline behavior with better accuracy at the training stage would likely improve the inference,
and we leave it as future work.
In addition, we might explore even more complex $\ppop(\vec{\theta}|\vec{\Lambda})$ parametrizations to capture with greater faithfulness the
astrophysical features of catalogs, which might be responsible for some of the observed differences between the inferences on MCMC fits and GPR predictions.

The present study is subject to other limitations, which can nevertheless be tackled. More in detail, we assumed a perfect subtraction of resolvable 
sources and perfect knowledge 
of the instrumental noise spectrum. In realistic LISA observations, uncertainties in the instrumental noise will inevitably introduce degeneracies 
with the confusion foreground, potentially impacting inference. Accounting for these effects will require a joint treatment of instrumental 
and astrophysical noise components within the inference framework. Furthermore, when subtracting the sources we assumed a hard threshold in SNR for
resolvabilty. This can be relaxed by moving to a probabilistic treatment of resolvability, which would account for some sources to remain
unresolved and therefore contribute to the foreground.

Several extensions of this work are therefore natural. Future developments will further include the impact of the yearly modulation induced by LISA’s orbital motion
into the foreground which might help in disentangling instrumental noise uncertainties.
Moreover, recent studies~\cite{Fabbri:2025faf, Rinaldi:2025evs} have explored the idea of mapping non-parametric reconstructions of population properties into parametric models.
Combining such approaches with the framework presented here may provide a promising route to connect agnostic population parametrizations with physically motivated descriptions. 
Future work will explore a broader range of astrophysical models to ensure that the adopted parametrization is sufficiently flexible 
to capture them, enabling a faithful representation of the GB population and allowing such a mapping procedure to be used for model selection.
Finally, we note that the methods presented in this work could, in principle, be exploited directly at the level of global-fit pipelines 
in terms of population-informed priors which might improve their robustness and overall performance.

\acknowledgements
We thank Tristan Bruel, Fabiola Cocchiararo, Jonathan Gair, Astrid Lamberts, and Rodrigo Tenorio for discussions.
F.D.S., A.T. and D.G are supported by 
ERC Starting Grant No.~945155--GWmining, 
Cariplo Foundation Grant No.~2021-0555, 
MUR PRIN Grant No.~2022-Z9X4XS, 
Italian-French University (UIF/UFI) Grant No.~2025-C3-386,
MUR Grant ``Progetto Dipartimenti di Eccellenza 2023-2027'' (BiCoQ),
and the ICSC National Research Centre funded by NextGenerationEU.  
A.S. is supported by the German Space Agency (DLR) with funding from the Bundesministerium für Wirtschaft und Klimaschutz, based on a decision of the Deutsche Bundestag (Project Ref. No. FKZ 50
OQ 2301).
A.T. and D.G. are supported by MUR Young Researchers Grant No. SOE2024-0000125.
D.G. is supported by MSCA Fellowship No.~101149270--ProtoBH.
NK was supported by the Hellenic Foundation for Research and Innovation (H.F.R.I.) under the 4th Call for HFRI research projects to support post-doctoral researchers (Project No.~28418).
Computational work was performed 
at CINECA with allocations through INFN and the University of Milano-Bicocca, 
at NVIDIA with allocations through the Academic Grant program, and on the Saraswati and Lakshmi clusters at the Max Planck Institute for Gravitational Physics in Potsdam.

\section*{Data Availability Statement}
We make the subtraction algorithm code publicly available at Ref.~\cite{Fast_LISA_Subtraction}.
The data supporting this work will be shared upon reasonable request to the authors.

\appendix

\section{Copulas}\label{sec: copulas}
In Sec.~\ref{sec: population model} we use copulas to model the correlation between $\f$ and $\fdot$; for previous applications of copulas in GW astronomy, see e.g.~\cite{Adamcewicz:2022hce,2025arXiv250818083T}.
Given two random variables $x$ and $y$ with marginal distributions 
$p_x(x)$ and $p_y(y)$, their joint distribution can be factorized as~\cite{sklar2009copulas}
\begin{equation}\label{eq: copula joint dist}
    p(x, y \mid \corr) = p_x(x)\, p_y(y)\, \mathcal{C}\bigl(u(x), v(y) \mid \corr\bigr),
\end{equation}
where
\begin{align}\label{eq: copula CDFs}
        u(x) &= \int_{x_\text{min}}^{x} p_x(x')\, dx',\\
        \label{eq: copula CDFs2}
        v(y) &= \int_{y_\text{min}}^{y} p_y(y')\, dy'
\end{align}
are the CDFs of the marginal distributions, and $\mathcal{C}(u, v \mid \corr)$ is the copula function, which depends on the parameter $\corr$. This parameter controls the strength of the correlation between $x$ and $y$.

\noindent Specifically, we use a Gaussian copula
\begin{align}\label{eq: gaussian copula}
   \mathcal{C}_\Sigma(u, v \mid \corr) 
   &= \Phi_\Sigma\bigl(\Phi^{-1}(u), \Phi^{-1}(v)\bigr), 
   \\ 
   \Sigma &= 
   \begin{pmatrix}
        1 & \corr \\
        \corr & 1
   \end{pmatrix}, \;\, \corr < 1,
   \label{eq: gaussian copula2}\end{align}
where $\Phi^{-1}$ denotes the inverse CDF of the standard normal distribution, and $\Phi_\Sigma$ is the CDF of the bivariate normal distribution with zero mean and covariance matrix~$\Sigma$.

In practice, generating correlated samples from two arbitrary marginal distributions proceeds as follows:
\begin{enumerate}
    \item Draw independent samples $x$ and $y$ from their respective marginal distributions.
    \item Map these samples to $u, v \sim \mathcal{U}(0,1)$ via the marginal CDFs; Eqs.~(\ref{eq: copula CDFs})--(\ref{eq: copula CDFs2}).
    \item Sample $\boldsymbol{z} = (z_1, z_2) \sim \mathcal{N}(0, \Sigma)$ and set $u' = \Phi(z_1)$, $v' = \Phi(z_2)$, so that $(u', v')$ 
    has the correlation structure specified by Eqs.~\eqref{eq: gaussian copula}--\eqref{eq: gaussian copula2}.
    \item Apply the inverse marginal CDFs [Eqs.~(\ref{eq: copula CDFs})--(\ref{eq: copula CDFs2})] to $(u', v')$ to obtain correlated samples $(x', y')$ with the original marginals.
\end{enumerate}

\begin{figure}
    \centering
    \includegraphics[width=\columnwidth]{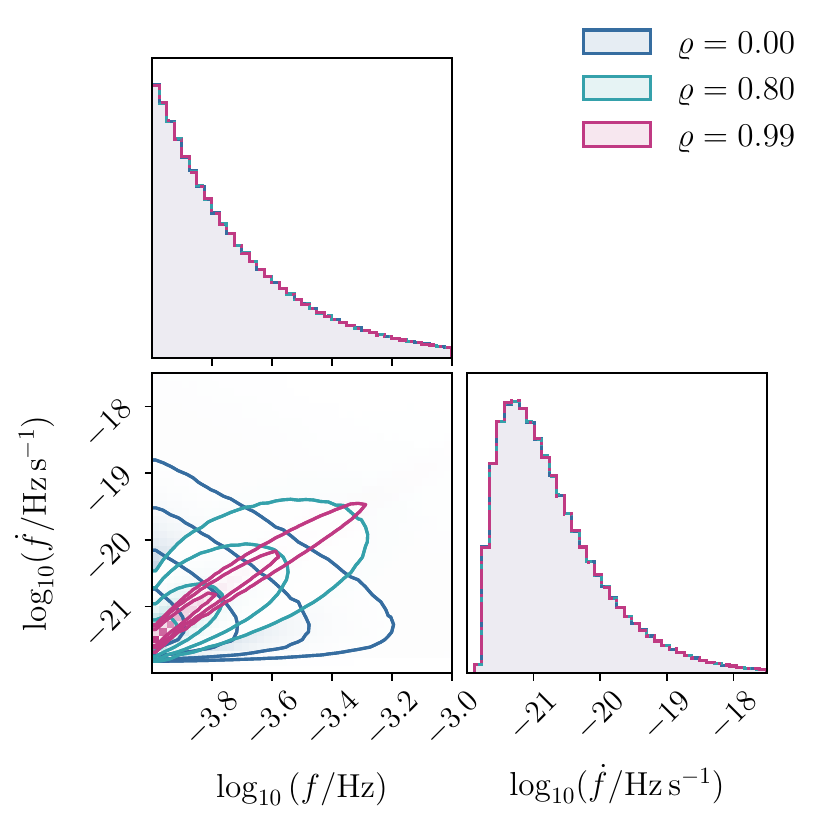}
    \caption{Example of the Gaussian copula applied to the joint distribution $p(\f, \fdot)$ for different values of $\corr$.}
    \label{fig: copula}
\end{figure}

We show an example of this procedure in Fig.~\ref{fig: copula}, where we plot the joint $f$--$\dot f$ distribution obtained for different 
values of the correlation coefficient $\corr$. Note that the marginal distributions are independent of the value of $\corr$.

\section{Characteristic strain}\label{sec: from PSD to characteristic strain}
The characteristic strain
\begin{equation}\label{eq: characteristic strain}
    h_c(f) = \sqrt{f\, S_n(f)}
\end{equation}
is commonly used to compare the sensitivity of a detector to gravitational waves emitted by astrophysical sources~\cite{Moore:2014lga},
where $S_n(f)$ denotes the detector power spectral density.

While Eq.~\eqref{eq: characteristic strain} is valid for ground-based detectors such as LIGO and Virgo, 
in the case of LISA one needs to rescale the noise covariance matrix $S_n^{ij}(f)$ by the TDI response 
$\mathcal{R}_{ij}(f)$~\cite{Flauger:2020qyi},
\begin{equation}\label{eq: characteristic strain2}
    S_{n}^{ij}(f) \rightarrow \frac{S_n^{ij}(f)}{\mathcal{R}_{ij}(f)} \, .
\end{equation}

The response functions $\mathcal{R}_{ij}$ are derived in Refs.~\cite{Flauger:2020qyi,Baghi:2023qnq} and take the form
\begin{equation}\label{eq: tdi response}
 \mathcal{R}_{ij}(f) = 16 \sin^2\!\left( \frac{f}{f^*} \right) 
 \left( \frac{f}{f^*} \right)^2 \tilde{\mathcal{R}}_{ij}(f),
\end{equation}
where $\tilde{\mathcal{R}}_{ij}(f)$ is approximately given by
\begin{align}
    \tilde{\mathcal{R}}_{\mathrm{AA}}(f) =  \tilde{\mathcal{R}}_{\mathrm{EE}}(f) &= 
    \frac{9}{20}\,\frac{1}{1 + 0.7 \left( \frac{f}{f^*} \right)^2}, 
    \label{eq: R tilde A}\\
    \tilde{\mathcal{R}}_{\mathrm{TT}}(f) &= 
    \frac{9}{20}\,
    \frac{\left( \frac{f}{f^*} \right)^6}{1.8 \times 10^3 + 0.7 \left( \frac{f}{f^*} \right)^8}.
    \label{eq: R tilde E}
\end{align}
\pagebreak

\section{LISA instrumental noise}\label{sec: lisa noise}

The LISA instrumental noise has two dominant contributions in the TDI variables~\cite{Flauger:2020qyi}: the optical
metrology system (OMS) noise and the test mass (TM) acceleration noise. 
The former includes shot-noise contributions, while the latter refers to the random displacement of the test masses. 
Current models for the noise spectra, based on the results of LISA Pathfinder~\cite{Armano:2016bkm} are given by
\begin{align}\label{eq: ims + acc noise}
        P_\text{OMS}(f) &= P^2 \frac{\SI{}{\pico\meter}^2}{\SI{}{\hertz}} 
        \left[ 1 + \left( \frac{\SI{2}{\milli\hertz}}{f} \right)^4\right]
        \left( \frac{2\pi f}{c}\right)^2 ,\\[1em]
        P_\text{TM}(f) &= A^2 \frac{\SI{}{\femto\meter}^2}{\SI{}{\second^4 \hertz}} 
        \left[ 1 + \left( \frac{\SI{0.4}{\milli\hertz}}{f} \right)^2\right] \notag \times \\
        & \times \left[1 + \left(\frac{f}{\SI{8}{\milli \hertz}^4} \right) \right] 
        \left( \frac{1}{2\pi f}\right)^4
        \left( \frac{2\pi f}{c}\right)^2 .
\end{align}

When assuming that the LISA constellation forms an equilateral triangle of fixed arms length $L$, 
the noise autocorrelation functions (i.e. the PSDs) for the TDI variables in the A-E-T basis are given by~\cite{Flauger:2020qyi}
\begin{align} 
& S_\text{instr}^\text{AA}(f) = S_\text{instr}^\text{EE}(f) =  8 \sin^2\!\left( \frac{f}{f^*}\right)  \! \left\lbrace
            4\left[ 1 + \cos\!\left( \frac{f}{f^*}\right) 
            \right.\right.
            \notag +\\
        & \quad \left.\left.
            + \cos^2\left( \frac{f}{f^*}\right) \right]
              P_\text{TM}(f)   
        + \left[2 + \cos\left( \frac{f}{f^*}\right) \right] P_\text{OMS}(f)
         \right\rbrace\,,
\\[1em]
     &   S_\text{instr}^\text{TT}(f) = 16 \sin^2\left( \frac{f}{f^*}\right) \left\lbrace
            2\left[ 1 - \cos\left( \frac{f}{f^*}\right) \right]^2    P_\text{TM}(f) 
            \right. \notag +\\
        &\quad \left.
        + \left[1 - \cos\left( \frac{f}{f^*}\right) \right] P_\text{OMS}(f)
        \right\rbrace\,,
    \end{align}
where we defined $f^* = c/(2\pi L)$ as the characteristic frequency of LISA. The current design sensitivity corresponds to $P = 15$ and $A = 3$ (with a $\pm 20\%$ margin), which is referred to as the \textsc{SciRDv1} noise budget~\cite{LISA_SciRDv1}.

Transfer functions for TDI 2.0 variables are derived in Ref.~\cite{QuangNam:2022gjz}. 
Switching to TDI 2.0 amounts to introducing an additional factor
\begin{equation}\label{eq: tdi2 conversion}
    S_{n, \AET}^\text{TDI2}(f) = 4 \sin^2 \left(2\frac{f}{f^*}\right) S_{n,\AET}(f)\,,
\end{equation}
which are the expressions used in this paper.

\section{Subtraction algorithm}\label{sec: subtraction algorithm}
\begin{figure}[!t]
  \centering
  \includegraphics[width=0.9\columnwidth]{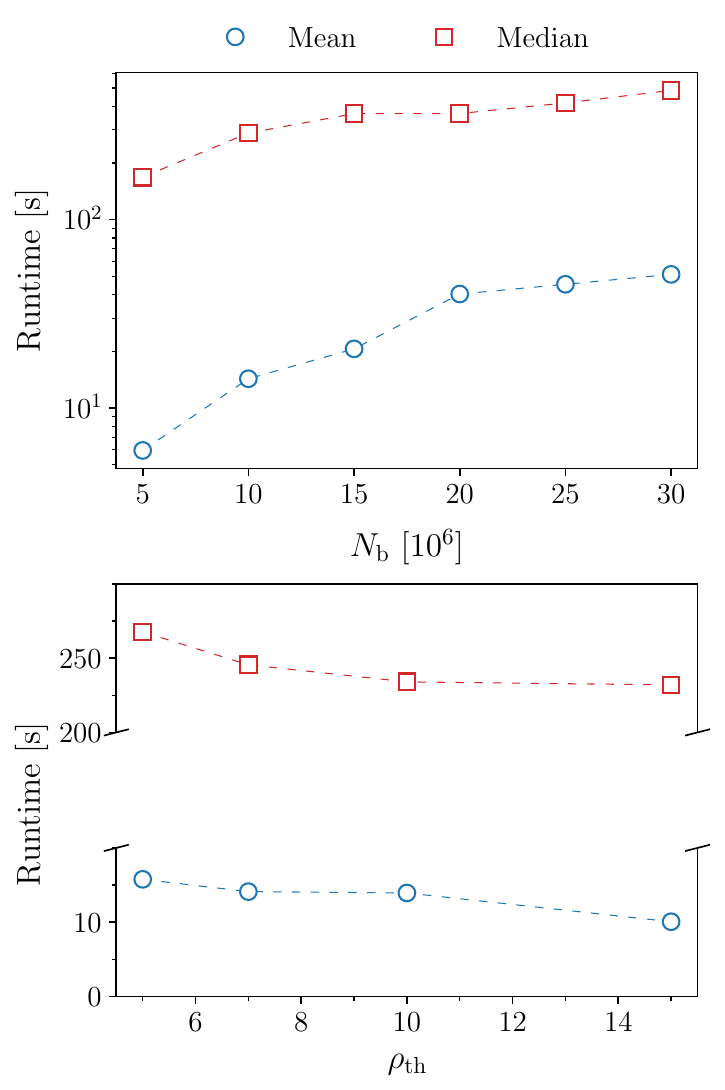}

  \caption{Execution time of the subtraction algorithm as a function of the number of sources in the catalog $\Nbinaries$ (top) and the
  SNR threshold $\rho_\text{th}$ (bottom) for the mean and median PSD estimators.
  Each point corresponds to the average of 20 runs, with catalogs randomly generated from the population model described in Sec.~\ref{sec: population model}.
  Computation is performed on a single NVIDIA RTX 4090 GPU.}
  \label{fig: subtraction time}
\end{figure}

The subtraction algorithm used in this paper was first introduced in Ref.~\cite{Karnesis:2021tsh} and consists of an iterative procedure to estimate the total 
power spectral density $\avgsconf = \sinstr + \sgal$ by subtracting, at each step, the contribution of binaries exceeding a given signal-to-noise ratio threshold $\rho_\text{th}$.
The main steps of the algorithm are summarized as follows: 

  \medskip
    \begin{algorithmic}[1]
    \STATE Generate a catalog of GBs using the population prior $\ppop(\vec{\theta} \mid \vec{\Lambda})$.
    \smallskip
    \STATE Compute the SNR $\rho$ of each binary using Eq.~\eqref{eq: SNR inner product} with $\sinstr$ as PSD.
    \smallskip
    \STATE Select binaries with $\rho \geq \kappa \,\rho_\text{th}$ ($\kappa < 1$)
    \smallskip
    \REPEAT
    \smallskip
        \STATE Estimate $S_n^{(i)}(f)$ with a running mean/median.
        \smallskip
        \STATE Compute the SNR $\rho$ of each binary using $S_n^{(i)}(f)$.
        \smallskip
        \FORALL{binaries with $\rho > \rho_\text{th}$}
        \smallskip
            \STATE Subtract their contribution from $h_\AET(f)$
            \smallskip
        \ENDFOR
        \smallskip
    \UNTIL{no binaries exceed the SNR threshold or convergence in $S_n^{(i)}(f)$ is reached.}
    \end{algorithmic}
    \medskip

The starting point is a frequency series containing the sum of all injected sources, $h_\AET=\sum_i h_{\AET, i}$, for each channel.
We fix the instrumental noise to the analytical \textsc{SciRDv1} model, $\sinstr$, and compute an initial estimate of the signal-to-noise ratio $\rho$ using Eq.~\eqref{eq: SNR inner product}.
We use this initial SNR estimate to select only sources with $\rho \geq 1$ (i.e. set $\kappa = 1/\rho_{\rm th})$, 
thereby reducing the computational load, since this cut removes sources that will definitely not be resolved during the subtraction procedure.

At each iteration the total PSD $S_n^{(i)}(f)$ is estimated, for each channel, with either a {running mean}
\begin{equation}\label{eq: running mean}
    S_n^{(i)}(f_k) = \frac{1}{N} \sum_{j=k-N/2}^{k+N/2} |2\Delta f \,h(f_j)|^2 + S_\text{instr}(f_j)
\end{equation}
or a {running median}
\begin{equation}\label{eq: running median}
    S_n^{(i)}(f_k) = \mathcal{C} \,\text{median}\left\{|2\Delta f \,h(f_j)|^2\right\}_{j=k-N/2}^{k+N/2} + S_\text{instr}(f_j)\,,
\end{equation}
\noindent where $N$ is the size of the window, $k$ runs over the frequency indexes. \\
The normalization constant $\mathcal{C}$ 
is given by
\begin{equation}
  \mathcal{C} = \left(1 - \frac{2}{9\nu}\right)^{-3} \approx 1.4237 \quad \text{for } \nu=2 
\end{equation}
since the PSD follows a $\chi^2_\nu$ distribution with $\nu=2$ degrees of freedom.
In this work, we set $N=2000$ and adopt the mean estimator of Eq.~\eqref{eq: running mean}.
The estimated PSD $S_n^{(i)}(f)$ is then used to recompute the SNR of each binary, and the contributions of all sources with 
$\rho > \rho_\text{th} = 7$ are subtracted from the data. This procedure is repeated until no binaries exceed the SNR threshold 
or convergence in $S_n^{(i)}$ is reached.

The subtraction algorithm is implemented in \textsc{Python} and accelerated on GPU using \textsc{CuPy}. 
Our implementation is available at  Ref.~\cite{Fast_LISA_Subtraction}.  
We exploit the GPU’s massive parallelism by batching all per-source operations: instead of looping over sources, 
we subtract the contributions of a batch of $\mathcal{O}(10^4)$ resolvable sources in a single vectorized step. 
Specifically, we use \texttt{cupy.add.at} to perform this operation on the frequency bins:
\begin{equation}\label{eq:vectorized-subtraction}
h_\AET[I_f] \leftarrow h_\AET[I_f] - \mathcal{H}_\AET,
\end{equation}
where $I_f \in \mathbb{N}^{N_{\text{batch}}\times N_f }$ contains, for each source (rows), the indices of the corresponding frequency bins 
in the data $h_\AET$, and $\mathcal{H}_\AET \in \mathbb{C}^{N_{\text{batch}}\times N_f }$ are the source templates to subtract.

Figure~\ref{fig: subtraction time} shows the execution runtime of the subtraction algorithm as a function of the number of 
sources in the catalog $\Nbinaries$ and the SNR threshold $\rho_\text{th}$ for both mean and median PSD estimators. 
The execution runtime scales approximately linearly with both $\Nbinaries$ and $\rho_\text{th}$, 
with the mean estimator being $\sim10\times$ faster than the median one.

\begin{table}[!t]
  \centering
  \renewcommand{\arraystretch}{1.5}
  \setlength{\tabcolsep}{6pt}
  \begin{tabular}{l cc cc}
   & \multicolumn{2}{c}{$N_\text{res}$} 
   & \multicolumn{2}{c}{Iterations} \\
   & Mean & Median
   & Mean & Median \\[1pt]
  \hline
  
  Ref.~\cite{Lamberts:2019nyk} 
   & 13102 & 14106 
   & 12 & 11 \\
  
  Simulation 
   & 31685 & 34854 
   & 17 & 15 \\
  \hline
  \end{tabular}
  \caption{Number of resolved sources $N_\text{res}$ and total iterations for the catalogs of Ref.~\cite{Lamberts:2019nyk} 
  and the simulation described in Sec.~\ref{sec: GPR}
  using mean and median estimators ($\rho_\text{th} = 7$).}
  \label{tab: mean median resolved}
  \end{table}

\section{Mean vs. median PSD estimators}\label{sec: mean vs median}
\begin{figure}[!t]
  \centering
  \includegraphics[width=0.9\columnwidth]{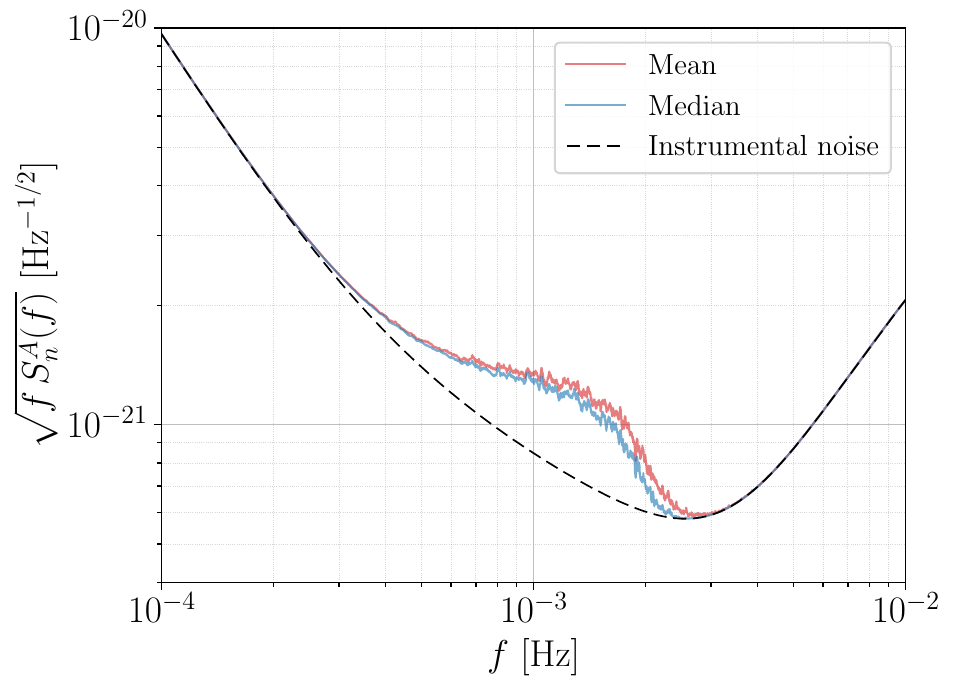}
  \includegraphics[width=0.9\columnwidth]{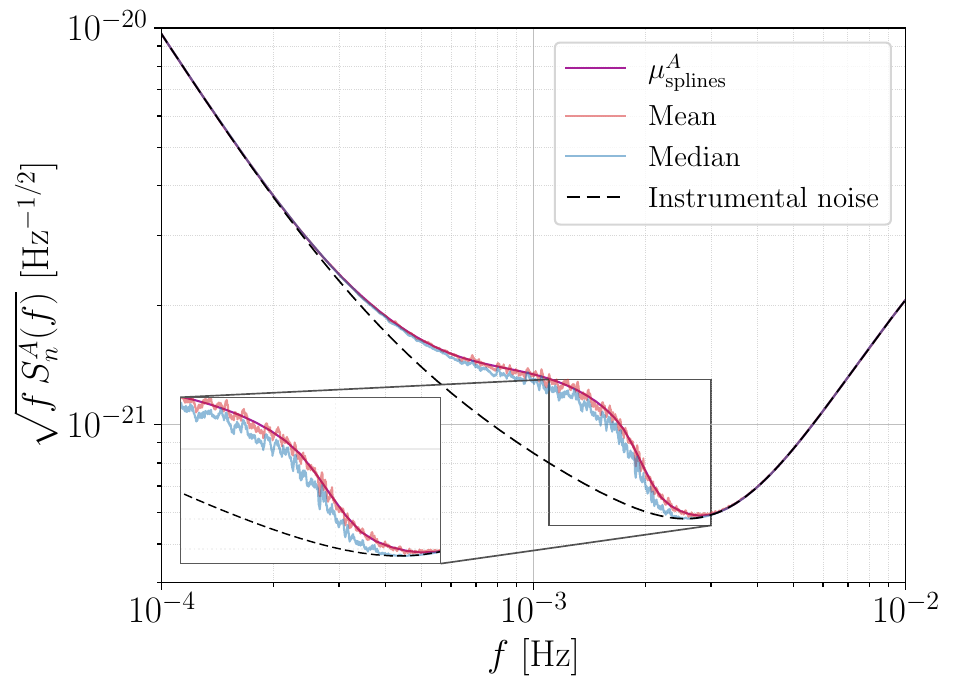}
  \caption{\textit{Top:} Total PSD $\sconf$ obtained by running the subtraction algorithm with 
  the mean and median estimators for the catalog of Ref.~\cite{Lamberts:2019nyk}.
  \textit{Bottom:} Total PSD $\sconf$ obtained by running the subtraction algorithm with the {median} estimator for the same catalog of Ref.~\cite{Lamberts:2019nyk}
  and then applying the mean and median estimators to the residual data $h_\AET$ after subtraction.  A spline fit to the data appears to be consistent with the mean estimator.
  }
  \label{fig: mean vs median}
\end{figure}

Figure~\ref{fig: mean vs median} compare results obtained with the running mean and median PSD estimators introduced in
Eqs.~\eqref{eq: running mean}--\eqref{eq: running median} considering the astrophysical catalog of Ref.~\cite{Lamberts:2019nyk}. Some related metrics are reported in
Table~\ref{tab: mean median resolved}, together with similar results obtained using the simulation described in Sec.~\ref{sec: GPR}.

The top panel of Fig.~\ref{fig: mean vs median} shows the outputs of the subtraction algorithm. We find the median estimator yields 
a lower foreground, especially at  frequencies $f \gtrsim 1$ mHz,
due to a higher number of resolved sources (cf. Table~\ref{tab: mean median resolved}).
These differences are likely due to the fact that the median estimator is more robust to outliers, which in this case correspond to the contribution of bright sources.

The bottom panel of Fig.~\ref{fig: mean vs median} shows instead a comparison with a spline fit to the data. 
For this particular test we ran the subtraction algorithm with the {median} 
estimator and then apply Eqs.~\eqref{eq: running mean}--\eqref{eq: running median} to the residual data $h_\AET$ 
after the subtraction. The median curves in the top and bottom panels of Fig.~\ref{fig: mean vs median} 
are indeed the same, 
whereas the mean curves are different, as they have been obtained by applying Eq.~\eqref{eq: running mean} 
to different residual data.
The spline fit adopted in the MCMC inference appears to be consistent with the mean estimator. 
Therefore, to maintain full consistency throughout the analysis pipeline and avoid additional biases, 
we adopt the mean estimator in the subtraction algorithm when simulating the data.

\bibliography{foreground}

@article{Tinto:2004wu,
	archiveprefix = {arXiv},
	author = {Tinto, Massimo and Dhurandhar, Sanjeev V.},
	doi = {10.12942/lrr-2005-4},
	eprint = {gr-qc/0409034},
	journal = {Living Rev. Relativ.},
	pages = {4},
	title = {{TIME DELAY}},
	volume = {8},
	year = {2005}}

@article{2026arXiv260104168T,
	adsurl = {https://ui.adsabs.harvard.edu/abs/2026arXiv260104168T},
	archiveprefix = {arXiv},
	author = {{Toubiana}, Alexandre and {Gair}, Jonathan},
	eprint = {2601.04168},
	journal = {{}},
	month = jan,
	primaryclass = {gr-qc},
	title = {{A framework for LISA population inference}},
	year = 2026}

@article{Strub:2024kbe,
	archiveprefix = {arXiv},
	author = {Strub, Stefan H. and Ferraioli, Luigi and Schmelzbach, C\'edric and St\"ahler, Simon C. and Giardini, Domenico},
	doi = {10.1103/PhysRevD.110.024005},
	eprint = {2403.15318},
	journal = {Phys. Rev. D},
	number = {2},
	pages = {024005},
	primaryclass = {gr-qc},
	title = {{Global analysis of LISA data with Galactic binaries and massive black hole binaries}},
	volume = {110},
	year = {2024}}

@article{Ruiter:2007xx,
	archiveprefix = {arXiv},
	author = {Ruiter, Ashley J. and Belczynski, Krzysztof and Benacquista, Matthew and Larson, Shane L. and Williams, Gabriel},
	doi = {10.1088/0004-637X/717/2/1006},
	eprint = {0705.3272},
	journal = {Astrophys. J.},
	pages = {1006--1021},
	primaryclass = {astro-ph},
	title = {{The LISA Gravitational Wave Foreground: A Study of Double White Dwarfs}},
	volume = {717},
	year = {2010}}

@article{Nelemans:2001hp,
	archiveprefix = {arXiv},
	author = {Nelemans, G. and Yungelson, L. R. and Portegies Zwart, Simon F.},
	doi = {10.1051/0004-6361:20010683},
	eprint = {astro-ph/0105221},
	journal = {Astron. Astrophys.},
	pages = {890--898},
	title = {{The gravitational wave signal from the galactic disk population of binaries containing two compact objects}},
	volume = {375},
	year = {2001}}

@article{Breivik:2017jip,
	archiveprefix = {arXiv},
	author = {Breivik, Katelyn and Kremer, Kyle and Bueno, Michael and Larson, Shane L. and Coughlin, Scott and Kalogera, Vassiliki},
	doi = {10.3847/2041-8213/aaaa23},
	eprint = {1710.08370},
	journal = {Astrophys. J. Lett.},
	number = {1},
	pages = {L1},
	primaryclass = {astro-ph.SR},
	title = {{Characterizing Accreting Double White Dwarf Binaries with the Laser Interferometer Space Antenna and Gaia}},
	volume = {854},
	year = {2018}}

@article{2002MNRAS.329..897H,
	adsurl = {https://ui.adsabs.harvard.edu/abs/2002MNRAS.329..897H},
	archiveprefix = {arXiv},
	author = {{Hurley}, Jarrod R. and {Tout}, Christopher A. and {Pols}, Onno R.},
	doi = {10.1046/j.1365-8711.2002.05038.x},
	eprint = {astro-ph/0201220},
	journal = {Mon. Not. R. Astron. Soc.},
	month = feb,
	number = {4},
	pages = {897-928},
	primaryclass = {astro-ph},
	title = {{Evolution of binary stars and the effect of tides on binary populations}},
	volume = {329},
	year = 2002}

@article{2024A&A...692A.165T,
	adsurl = {https://ui.adsabs.harvard.edu/abs/2024A&A...692A.165T},
	archiveprefix = {arXiv},
	author = {{Toubiana}, A. and {Karnesis}, N. and {Lamberts}, A. and {Miller}, M.~C.},
	doi = {10.1051/0004-6361/202450174},
	eid = {A165},
	eprint = {2403.16867},
	journal = {Astron. Astrophys.},
	month = dec,
	pages = {A165},
	primaryclass = {astro-ph.SR},
	title = {{The interacting double white dwarf population with LISA: Stochastic foreground and resolved sources}},
	volume = {692},
	year = 2024}

@article{Littenberg:2020bxy,
	author = {Littenberg, Tyson B. and Cornish, Neil J. and Lackeos, Kristen and Robson, Travis},
	doi = {10.1103/PhysRevD.101.123021},
	issue = {12},
	journal = {Phys. Rev. D},
	month = {Jun},
	numpages = {17},
	pages = {123021},
	publisher = {American Physical Society},
	title = {Global analysis of the gravitational wave signal from Galactic binaries},
	url = {https://link.aps.org/doi/10.1103/PhysRevD.101.123021},
	volume = {101},
	year = {2020}}

@article{Korol:2021pun,
	archiveprefix = {arXiv},
	author = {Korol, Valeriya and Hallakoun, Na'ama and Toonen, Silvia and Karnesis, Nikolaos},
	doi = {10.1093/mnras/stac415},
	eprint = {2109.10972},
	journal = {Mon. Not. R. Astron. Soc.},
	number = {4},
	pages = {5936--5947},
	primaryclass = {astro-ph.HE},
	title = {{Observationally driven Galactic double white dwarf population for LISA}},
	volume = {511},
	year = {2022}}

@article{Karnesis:2021tsh,
	archiveprefix = {arXiv},
	author = {Karnesis, Nikolaos and Babak, Stanislav and Pieroni, Mauro and Cornish, Neil and Littenberg, Tyson},
	doi = {10.1103/PhysRevD.104.043019},
	eprint = {2103.14598},
	journal = {Phys. Rev. D},
	number = {4},
	pages = {043019},
	primaryclass = {astro-ph.IM},
	title = {{Characterization of the stochastic signal originating from compact binary populations as measured by LISA}},
	volume = {104},
	year = {2021}}

@article{GPflow2017,
	archiveprefix = {arXiv},
	author = {Matthews, Alexander G. de G. and {van der Wilk}, Mark and Nickson, Tom and Fujii, Keisuke. and {Boukouvalas}, Alexis and {Le{\'o}n-Villagr{\'a}}, Pablo and Ghahramani, Zoubin and Hensman, James},
	eprint = {1610.08733},
	journal = {J. Mach. Learn. Res.},
	month = {apr},
	number = {40},
	pages = {1-6},
	primaryclass = {stat.ML},
	title = {{{GP}flow: A {G}aussian process library using {T}ensor{F}low}},
	url = {http://jmlr.org/papers/v18/16-537.html},
	volume = {18},
	year = {2017}}

@article{GPflow2020multioutput,
	author = {{van der Wilk}, Mark and Dutordoir, Vincent and John, ST and Artemev, Artem and Adam, Vincent and Hensman, James},
	journal = {arXiv:2003.01115},
	title = {A Framework for Interdomain and Multioutput {G}aussian Processes},
	url = {https://arxiv.org/abs/2003.01115},
	year = {2020}}

@article{Caprini:2019pxz,
	archiveprefix = {arXiv},
	author = {Caprini, Chiara and Figueroa, Daniel G. and Flauger, Raphael and Nardini, Germano and Peloso, Marco and Pieroni, Mauro and Ricciardone, Angelo and Tasinato, Gianmassimo},
	doi = {10.1088/1475-7516/2019/11/017},
	eprint = {1906.09244},
	journal = {J. Cosmology Astropart. Phys.},
	pages = {017},
	primaryclass = {astro-ph.CO},
	reportnumber = {LISA-CosWG-19-02},
	title = {{Reconstructing the spectral shape of a stochastic gravitational wave background with LISA}},
	volume = {11},
	year = {2019}}

@article{Flauger:2020qyi,
	archiveprefix = {arXiv},
	author = {Flauger, Raphael and Karnesis, Nikolaos and Nardini, Germano and Pieroni, Mauro and Ricciardone, Angelo and Torrado, Jes\'us},
	doi = {10.1088/1475-7516/2021/01/059},
	eprint = {2009.11845},
	journal = {J. Cosmology Astropart. Phys.},
	pages = {059},
	primaryclass = {astro-ph.CO},
	title = {{Improved reconstruction of a stochastic gravitational wave background with LISA}},
	volume = {01},
	year = {2021}}

@article{QuangNam:2022gjz,
	archiveprefix = {arXiv},
	author = {Quang Nam, Dam and Lemi\`ere, Yves and Petiteau, Antoine and Bayle, Jean-Baptiste and Hartwig, Olaf and Martino, Joseph and Staab, Martin},
	doi = {10.1103/PhysRevD.108.082004},
	eprint = {2211.02539},
	journal = {Phys. Rev. D},
	number = {8},
	pages = {082004},
	primaryclass = {gr-qc},
	title = {{Time-delay interferometry noise transfer functions for LISA}},
	volume = {108},
	year = {2023}}

@article{Katz:2024oqg,
	archiveprefix = {arXiv},
	author = {Katz, Michael L. and Karnesis, Nikolaos and Korsakova, Natalia and Gair, Jonathan R. and Stergioulas, Nikolaos},
	doi = {10.1103/PhysRevD.111.024060},
	eprint = {2405.04690},
	journal = {Phys. Rev. D},
	number = {2},
	pages = {024060},
	primaryclass = {gr-qc},
	title = {{Efficient GPU-accelerated multisource global fit pipeline for LISA data analysis}},
	volume = {111},
	year = {2025}}

@article{Deng:2025wgk,
	archiveprefix = {arXiv},
	author = {Deng, Senwen and Babak, Stanislav and Le Jeune, Maude and Marsat, Sylvain and Plagnol, {\'E}ric and Sartirana, Andrea},
	doi = {10.1103/PhysRevD.111.103014},
	eprint = {2501.10277},
	journal = {Phys. Rev. D},
	number = {10},
	pages = {103014},
	primaryclass = {gr-qc},
	title = {{Modular global-fit pipeline for LISA data analysis}},
	volume = {111},
	year = {2025}}

@article{Lamberts:2019nyk,
	archiveprefix = {arXiv},
	author = {Lamberts, Astrid and Blunt, Sarah and Littenberg, Tyson B. and Garrison-Kimmel, Shea and Kupfer, Thomas and Sanderson, Robyn E.},
	doi = {10.1093/mnras/stz2834},
	eprint = {1907.00014},
	journal = {Mon. Not. R. Astron. Soc.},
	number = {4},
	pages = {5888--5903},
	primaryclass = {astro-ph.HE},
	title = {{Predicting the LISA white dwarf binary population in the Milky Way with cosmological simulations}},
	volume = {490},
	year = {2019}}

@article{Korol:2017qcx,
	archiveprefix = {arXiv},
	author = {Korol, Valeriya and Rossi, Elena M. and Groot, Paul J. and Nelemans, Gijs and Toonen, Silvia and Brown, Anthony G. A.},
	doi = {10.1093/mnras/stx1285},
	eprint = {1703.02555},
	journal = {Mon. Not. R. Astron. Soc.},
	number = {2},
	pages = {1894--1910},
	primaryclass = {astro-ph.HE},
	title = {{Prospects for detection of detached double white dwarf binaries with Gaia, LSST and LISA}},
	volume = {470},
	year = {2017}}

@article{Baghi:2023qnq,
	archiveprefix = {arXiv},
	author = {Baghi, Quentin and Karnesis, Nikolaos and Bayle, Jean-Baptiste and Besan{\c{c}}on, Marc and Inchausp{\'e}, Henri},
	doi = {10.1088/1475-7516/2023/04/066},
	eprint = {2302.12573},
	journal = {J. Cosmology Astropart. Phys.},
	pages = {066},
	primaryclass = {gr-qc},
	title = {{Uncovering gravitational-wave backgrounds from noises of unknown shape with LISA}},
	volume = {04},
	year = {2023}}

@article{Maoz:2018epf,
	archiveprefix = {arXiv},
	author = {Maoz, Dan and Hallakoun, Na'ama and Badenes, Carles},
	doi = {10.1093/mnras/sty339},
	eprint = {1801.04275},
	journal = {Mon. Not. R. Astron. Soc.},
	number = {2},
	pages = {2584--2590},
	primaryclass = {astro-ph.SR},
	title = {{The separation distribution and merger rate of double white dwarfs: improved constraints}},
	volume = {476},
	year = {2018}}

@article{Maoz:2012aj,
	archiveprefix = {arXiv},
	author = {Maoz, Dan and Badenes, Carles and Bickerton, Steven J.},
	doi = {10.1088/0004-637X/751/2/143},
	eprint = {1202.5467},
	journal = {Astrophys. J.},
	pages = {143},
	primaryclass = {astro-ph.SR},
	title = {{Characterizing the Galactic White Dwarf Binary Population with Sparsely Sampled Radial Velocity Data}},
	volume = {751},
	year = {2012}}

@article{Maoz:2016bxg,
	archiveprefix = {arXiv},
	author = {Maoz, Dan and Hallakoun, Na'ama},
	doi = {10.1093/mnras/stx102},
	eprint = {1609.02156},
	journal = {Mon. Not. R. Astron. Soc.},
	number = {2},
	pages = {1414--1425},
	primaryclass = {astro-ph.SR},
	title = {{The binary fraction, separation distribution and merger rate of white dwarfs from SPY}},
	volume = {467},
	year = {2017}}

@article{Georgousi:2022uyt,
	archiveprefix = {arXiv},
	author = {Georgousi, Maria and Karnesis, Nikolaos and Korol, Valeriya and Pieroni, Mauro and Stergioulas, Nikolaos},
	doi = {10.1093/mnras/stac3686},
	eprint = {2204.07349},
	journal = {Mon. Not. R. Astron. Soc.},
	number = {2},
	pages = {2552--2566},
	primaryclass = {astro-ph.GA},
	title = {{Gravitational waves from double white dwarfs as probes of the milky way}},
	volume = {519},
	year = {2022}}

@article{Pieroni:2020rob,
	archiveprefix = {arXiv},
	author = {Pieroni, Mauro and Barausse, Enrico},
	doi = {10.1088/1475-7516/2020/07/021},
	eprint = {2004.01135},
	journal = {J. Cosmology Astropart. Phys.},
	note = {[Erratum: J. Cosmology Astropart. Phys. 09, E01 (2020)]},
	pages = {021},
	primaryclass = {astro-ph.CO},
	title = {{Foreground cleaning and template-free stochastic background extraction for LISA}},
	volume = {07},
	year = {2020}}

@article{Fabbri:2025faf,
	archiveprefix = {arXiv},
	author = {Fabbri, Cecilia Maria and Gerosa, Davide and Santini, Alessandro and Mould, Matthew and Toubiana, Alexandre and Gair, Jonathan},
	doi = {10.1103/PhysRevD.111.104053},
	eprint = {2501.17233},
	journal = {Phys. Rev. D},
	number = {10},
	pages = {104053},
	primaryclass = {astro-ph.HE},
	title = {{Reconstructing parametric gravitational-wave population fits from nonparametric results without refitting the data}},
	volume = {111},
	year = {2025}}

@article{Adamcewicz:2022hce,
	archiveprefix = {arXiv},
	author = {Adamcewicz, Christian and Thrane, Eric},
	doi = {10.1093/mnras/stac2961},
	eprint = {2208.03405},
	journal = {Mon. Not. R. Astron. Soc.},
	number = {3},
	pages = {3928--3937},
	primaryclass = {astro-ph.HE},
	title = {{Do unequal-mass binary black hole systems have larger {\ensuremath{\chi}}eff? Probing correlations with copulas in gravitational-wave astronomy}},
	volume = {517},
	year = {2022}}

@article{DeSanti:2024oap,
	author = {De Santi, Federico and Razzano, Massimiliano and Fidecaro, Francesco and Muccillo, Luca and Papalini, Lucia and Patricelli, Barbara},
	doi = {10.1103/PhysRevD.109.102004},
	issue = {10},
	journal = {Phys. Rev. D},
	month = {May},
	numpages = {21},
	pages = {102004},
	publisher = {American Physical Society},
	title = {Deep learning to detect gravitational waves from binary close encounters: Fast parameter estimation using normalizing flows},
	url = {https://link.aps.org/doi/10.1103/PhysRevD.109.102004},
	volume = {109},
	year = {2024}}

@article{Santini:2025iuj,
	archiveprefix = {arXiv},
	author = {Santini, Alessandro and Muratore, Martina and Gair, Jonathan and Hartwig, Olaf},
	doi = {10.1103/csx9-9trp},
	eprint = {2507.06300},
	journal = {Phys. Rev. D},
	number = {8},
	pages = {084050},
	primaryclass = {gr-qc},
	title = {{Flexible, GPU-accelerated approach for the joint characterization of LISA instrumental noise and stochastic gravitational wave backgrounds}},
	volume = {112},
	year = {2025}}

@article{GBGPU_michael_l_katz_2022_6500434,
	author = {Michael L. Katz},
	journal = {\href{https://doi.org/10.5281/zenodo.6500434}{doi.org/10.5281/zenodo.6500434}, \href{https://github.com/mikekatz04/gbgpu}{github.com/mikekatz04/gbgpu}},
	year = 2025}

@article{Katz:2022izt,
	archiveprefix = {arXiv},
	author = {Katz, Michael L. and Danielski, Camilla and Karnesis, Nikolaos and Korol, Valeriya and Tamanini, Nicola and Cornish, Neil J. and Littenberg, Tyson B.},
	doi = {10.1093/mnras/stac2555},
	eprint = {2205.03461},
	journal = {Mon. Not. R. Astron. Soc.},
	number = {1},
	pages = {697--711},
	primaryclass = {astro-ph.EP},
	title = {{Bayesian characterization of circumbinary sub-stellar objects with LISA}},
	volume = {517},
	year = {2022}}

@article{Cornish:2007if,
	archiveprefix = {arXiv},
	author = {Cornish, Neil J. and Littenberg, Tyson B.},
	doi = {10.1103/PhysRevD.76.083006},
	eprint = {0704.1808},
	journal = {Phys. Rev. D},
	pages = {083006},
	primaryclass = {gr-qc},
	title = {{Tests of Bayesian Model Selection Techniques for Gravitational Wave Astronomy}},
	volume = {76},
	year = {2007}}

@article{HYPERION,
	author = {{De Santi}, Federico},
	journal = {\href{https://github.com/fdesanti/hyperion}{github.com/fdesanti/hyperion}},
	year = {2024}}

@article{Korol:2020hay,
	archiveprefix = {arXiv},
	author = {Korol, Valeriya and Belokurov, Vasily and Moore, Christopher J. and Toonen, Silvia},
	doi = {10.1093/mnrasl/slab003},
	eprint = {2010.05918},
	journal = {Mon. Not. R. Astron. Soc.},
	number = {1},
	pages = {L55--L60},
	primaryclass = {astro-ph.GA},
	title = {{Weighing Milky Way Satellites with LISA}},
	volume = {502},
	year = {2021}}

@article{sklar2009copulas,
	author = {Abe Sklar},
	doi = {10.1214/lnms/1215452606},
	journal = {IMS Lecture Notes Monogr. Ser.},
	pages = {1--14},
	title = {Random Variables, Distribution Functions, and Copulas -- A Personal Look Back},
	url = {https://doi.org/10.1214/lnms/1215452606},
	volume = {1},
	year = {2009}}

@article{Robson:2018ifk,
	archiveprefix = {arXiv},
	author = {Robson, Travis and Cornish, Neil J. and Liu, Chang},
	doi = {10.1088/1361-6382/ab1101},
	eprint = {1803.01944},
	journal = {Class. Quantum Grav.},
	number = {10},
	pages = {105011},
	primaryclass = {astro-ph.HE},
	title = {{The construction and use of LISA sensitivity curves}},
	volume = {36},
	year = {2019}}

@article{Delfavero:2024zyl,
	archiveprefix = {arXiv},
	author = {Delfavero, Vera and Breivik, Katelyn and Thiele, Sarah and O'Shaughnessy, Richard and Baker, John G.},
	doi = {10.3847/1538-4357/ada9e2},
	eprint = {2409.15230},
	journal = {Astrophys. J.},
	number = {1},
	pages = {66},
	primaryclass = {gr-qc},
	title = {{Recovering Injected Astrophysics from the LISA Double White Dwarf Binaries}},
	volume = {981},
	year = {2025}}

@article{Srinivasan:2025etu,
	archiveprefix = {arXiv},
	author = {Srinivasan, Rahul and Barausse, Enrico and Korsakova, Natalia and Trotta, Roberto},
	doi = {10.1103/shym-w46f},
	eprint = {2506.22543},
	journal = {Phys. Rev. D},
	number = {10},
	pages = {103043},
	primaryclass = {astro-ph.GA},
	title = {{Simulation-based population inference of LISA{\textquoteright}s Galactic binaries: Bypassing the global fit}},
	volume = {112},
	year = {2025}}

@article{Hartwig:2023pft,
	archiveprefix = {arXiv},
	author = {Hartwig, Olaf and Lilley, Marc and Muratore, Martina and Pieroni, Mauro},
	doi = {10.1103/PhysRevD.107.123531},
	eprint = {2303.15929},
	journal = {Phys. Rev. D},
	number = {12},
	pages = {123531},
	primaryclass = {gr-qc},
	reportnumber = {CERN-TH-2023-050},
	title = {{Stochastic gravitational wave background reconstruction for a nonequilateral and unequal-noise LISA constellation}},
	volume = {107},
	year = {2023}}

@article{Armano:2016bkm,
	author = {Armano, M. and others},
	doi = {10.1103/PhysRevLett.116.231101},
	journal = {Phys. Rev. Lett.},
	number = {23},
	pages = {231101},
	title = {{Sub-Femto- g Free Fall for Space-Based Gravitational Wave Observatories: LISA Pathfinder Results}},
	volume = {116},
	year = {2016}}

@article{LISA_SciRDV1,
	archiveprefix = {arXiv},
	author = {{LISA Science Study Team}},
	journal = {ESA-L3-EST-SCI-RS001 \href{https://www.cosmos.esa.int/documents/678316/1700384/SciRD.pdf}{www.cosmos.esa.int/documents/678316/1700384/SciRD.pdf}},
	year = {2018}}

@article{2015ApJ...805L...6S,
	adsurl = {https://ui.adsabs.harvard.edu/abs/2015ApJ...805L...6S},
	archiveprefix = {arXiv},
	author = {{Shen}, Ken J.},
	doi = {10.1088/2041-8205/805/1/L6},
	eid = {L6},
	eprint = {1502.05052},
	journal = {Astrophys. J. Lett.},
	month = may,
	number = {1},
	pages = {L6},
	primaryclass = {astro-ph.SR},
	title = {{Every Interacting Double White Dwarf Binary May Merge}},
	volume = {805},
	year = 2015}

@article{Maoz:2013hna,
	archiveprefix = {arXiv},
	author = {Maoz, Dan and Mannucci, Filippo and Nelemans, Gijs},
	doi = {10.1146/annurev-astro-082812-141031},
	eprint = {1312.0628},
	journal = {Annu. Rev. Astron. Astrophys.},
	pages = {107--170},
	primaryclass = {astro-ph.CO},
	title = {{Observational clues to the progenitors of Type-Ia supernovae}},
	volume = {52},
	year = {2014}}

@article{Shen:2017flp,
	archiveprefix = {arXiv},
	author = {Shen, Ken J. and Kasen, Daniel and Miles, Broxton J. and Townsley, Dean M.},
	doi = {10.3847/1538-4357/aaa8de},
	eprint = {1706.01898},
	journal = {Astrophys. J.},
	number = {1},
	pages = {52},
	primaryclass = {astro-ph.HE},
	title = {{Sub-Chandrasekhar-mass white dwarf detonations revisited}},
	volume = {854},
	year = {2018}}

@article{2015ApJ...806...76K,
	adsurl = {https://ui.adsabs.harvard.edu/abs/2015ApJ...806...76K},
	archiveprefix = {arXiv},
	author = {{Kremer}, Kyle and {Sepinsky}, Jeremy and {Kalogera}, Vassiliki},
	doi = {10.1088/0004-637X/806/1/76},
	eid = {76},
	eprint = {1502.06147},
	journal = {Astrophys. J.},
	month = jun,
	number = {1},
	pages = {76},
	primaryclass = {astro-ph.SR},
	title = {{Long-term Evolution of Double White Dwarf Binaries Accreting through Direct Impact}},
	volume = {806},
	year = 2015}

@article{Moore:2014lga,
	archiveprefix = {arXiv},
	author = {Moore, C. J. and Cole, R. H. and Berry, C. P. L.},
	doi = {10.1088/0264-9381/32/1/015014},
	eprint = {1408.0740},
	journal = {Class. Quantum Grav.},
	number = {1},
	pages = {015014},
	primaryclass = {gr-qc},
	reportnumber = {LIGO-P1400129},
	title = {{Gravitational-wave sensitivity curves}},
	volume = {32},
	year = {2015}}

@article{2004MNRAS.350..113M,
	adsurl = {https://ui.adsabs.harvard.edu/abs/2004MNRAS.350..113M},
	archiveprefix = {arXiv},
	author = {{Marsh}, T.~R. and {Nelemans}, G. and {Steeghs}, D.},
	doi = {10.1111/j.1365-2966.2004.07564.x},
	eprint = {astro-ph/0312577},
	journal = {Mon. Not. R. Astron. Soc.},
	month = may,
	number = {1},
	pages = {113-128},
	primaryclass = {astro-ph},
	title = {{Mass transfer between double white dwarfs}},
	volume = {350},
	year = 2004}

@article{2021ApJ...908....1S,
	adsurl = {https://ui.adsabs.harvard.edu/abs/2021ApJ...908....1S},
	archiveprefix = {arXiv},
	author = {{Sberna}, Laura and {Toubiana}, Alexandre and {Miller}, M. Coleman},
	doi = {10.3847/1538-4357/abccc7},
	eid = {1},
	eprint = {2010.05974},
	journal = {Astrophys. J.},
	month = feb,
	number = {1},
	pages = {1},
	primaryclass = {astro-ph.SR},
	title = {{Golden Galactic Binaries for LISA: Mass-transferring White Dwarf Black Hole Binaries}},
	volume = {908},
	year = 2021}

@article{2010PASP..122.1133S,
	adsurl = {https://ui.adsabs.harvard.edu/abs/2010PASP..122.1133S},
	author = {{Solheim}, J. -E.},
	doi = {10.1086/656680},
	journal = {Publ. Astron. Soc. Pac.},
	month = oct,
	number = {896},
	pages = {1133},
	title = {{AM CVn Stars: Status and Challenges}},
	volume = {122},
	year = 2010}

@article{Pozzoli:2024wfe,
	archiveprefix = {arXiv},
	author = {Pozzoli, Federico and Buscicchio, Riccardo and Klein, Antoine and Korol, Valeriya and Sesana, Alberto and Haardt, Francesco},
	doi = {10.1103/PhysRevD.111.063005},
	eprint = {2410.08274},
	journal = {Phys. Rev. D},
	number = {6},
	pages = {063005},
	primaryclass = {astro-ph.GA},
	title = {{Cyclostationary signals in LISA: A practical application to Milky~Way satellites}},
	volume = {111},
	year = {2025}}

@article{Dax:2021tsq,
	archiveprefix = {arXiv},
	author = {Dax, Maximilian and Green, Stephen R. and Gair, Jonathan and Macke, Jakob H. and Buonanno, Alessandra and Sch{\"o}lkopf, Bernhard},
	doi = {10.1103/PhysRevLett.127.241103},
	eprint = {2106.12594},
	journal = {Phys. Rev. Lett.},
	number = {24},
	pages = {241103},
	primaryclass = {gr-qc},
	reportnumber = {LIGO-P2100223},
	title = {{Real-Time Gravitational Wave Science with Neural Posterior Estimation}},
	volume = {127},
	year = {2021}}

@article{Alvey:2023npw,
	archiveprefix = {arXiv},
	author = {Alvey, James and Bhardwaj, Uddipta and Domcke, Valerie and Pieroni, Mauro and Weniger, Christoph},
	doi = {10.1103/PhysRevD.109.083008},
	eprint = {2309.07954},
	journal = {Phys. Rev. D},
	number = {8},
	pages = {083008},
	primaryclass = {gr-qc},
	reportnumber = {CERN-TH-2023-167},
	title = {{Simulation-based inference for stochastic gravitational wave background data analysis}},
	volume = {109},
	year = {2024}}

@article{2020PNAS..11730055C,
	adsurl = {https://ui.adsabs.harvard.edu/abs/2020PNAS..11730055C},
	archiveprefix = {arXiv},
	author = {{Cranmer}, Kyle and {Brehmer}, Johann and {Louppe}, Gilles},
	doi = {10.1073/pnas.1912789117},
	eprint = {1911.01429},
	journal = {Proc. Natl. Acad. Sci. USA},
	month = dec,
	number = {48},
	pages = {30055-30062},
	primaryclass = {stat.ML},
	title = {{The frontier of simulation-based inference}},
	volume = {117},
	year = 2020}

@article{2019arXiv191202762P,
	adsurl = {https://ui.adsabs.harvard.edu/abs/2019arXiv191202762P},
	archiveprefix = {arXiv},
	author = {{Papamakarios}, George and {Nalisnick}, Eric and {Jimenez Rezende}, Danilo and {Mohamed}, Shakir and {Lakshminarayanan}, Balaji},
	eprint = {1912.02762},
	journal = {J. Mach. Learn. Res.},
	month = dec,
	number = {57},
	pages = {1-64},
	primaryclass = {stat.ML},
	title = {{Normalizing Flows for Probabilistic Modeling and Inference}},
	url = {https://jmlr.org/papers/v22/19-1028.html},
	volume = {22},
	year = 2021}

@article{Wetzel:2016wro,
	archiveprefix = {arXiv},
	author = {Wetzel, Andrew R. and Hopkins, Philip F. and Kim, Ji-hoon and Faucher-Giguere, Claude-Andre and Keres, Dusan and Quataert, Eliot},
	doi = {10.3847/2041-8205/827/2/L23},
	eprint = {1602.05957},
	journal = {Astrophys. J. Lett.},
	number = {2},
	pages = {L23},
	primaryclass = {astro-ph.GA},
	title = {{Reconciling dwarf galaxies with $\Lambda$CDM cosmology: Simulating a realistic population of satellites around a Milky Way-mass galaxy}},
	volume = {827},
	year = {2016}}

@article{Fast_LISA_Subtraction,
	author = {{De Santi}, Federico and {Karnesis}, Nikolaos},
	journal = {\href{https://doi.org/10.5281/zenodo.18710899}{doi.org/10.5281/zenodo.18710899}, \href{https://github.com/fdesanti/fast-lisa-subtraction}{Fast LISA Subtraction}},
	year = 2026}

@article{McKay:1979,
	author = {M. D. McKay and R. J. Beckman and W. J. Conover},
	issn = {00401706},
	journal = {Technometrics},
	number = {2},
	pages = {239--245},
	publisher = {[Taylor & Francis, Ltd., American Statistical Association, American Society for Quality]},
	title = {A Comparison of Three Methods for Selecting Values of Input Variables in the Analysis of Output from a Computer Code},
	url = {http://www.jstor.org/stable/1268522},
	urldate = {2025-09-25},
	volume = {21},
	year = {1979}}

@article{2025arXiv250818083T,
	adsurl = {https://ui.adsabs.harvard.edu/abs/2025arXiv250818083T},
	archiveprefix = {arXiv},
	author = {Abac, A. G. and others},
	eprint = {2508.18083},
	journal = {{}},
	month = aug,
	primaryclass = {astro-ph.HE},
	title = {{GWTC-4.0: Population Properties of Merging Compact Binaries}},
	year = 2025}

@software{cupy,
	author = {Preferred Networks, Inc. and CuPy Development Team},
	date = {2025-08-18},
	note = {Accessed: 2025-09-26},
	title = {CuPy: NumPy \& SciPy for GPU-accelerated computing},
	url = {https://docs.cupy.dev/},
	version = {13.6.0}}

@article{Taylor:2018iat,
	archiveprefix = {arXiv},
	author = {Taylor, Stephen R. and Gerosa, Davide},
	doi = {10.1103/PhysRevD.98.083017},
	eprint = {1806.08365},
	journal = {Phys. Rev. D},
	number = {8},
	pages = {083017},
	primaryclass = {astro-ph.HE},
	title = {{Mining Gravitational-wave Catalogs To Understand Binary Stellar Evolution: A New Hierarchical Bayesian Framework}},
	volume = {98},
	year = {2018}}

@article{2016arXiv160508803D,
	adsurl = {https://ui.adsabs.harvard.edu/abs/2016arXiv160508803D},
	archiveprefix = {arXiv},
	author = {{Dinh}, Laurent and {Sohl-Dickstein}, Jascha and {Bengio}, Samy},
	eprint = {1605.08803},
	journal = {{International Conference on Learning Representations}},
	month = may,
	primaryclass = {cs.LG},
	title = {{Density estimation using Real NVP}},
	year = 2016}

@article{Adam_optimizer,
	archiveprefix = {arXiv},
	author = {Kingma, Diederik P. and Ba, Jimmy},
	eprint = {1412.6980},
	journal = {International Conference for Learning Representations},
	primaryclass = {cs.LG},
	title = {{Adam: A Method for Stochastic Optimization}},
	year = {2017}}

@article{Karnesis:2023ras,
	archiveprefix = {arXiv},
	author = {Karnesis, Nikolaos and Katz, Michael L. and Korsakova, Natalia and Gair, Jonathan R. and Stergioulas, Nikolaos},
	doi = {10.1093/mnras/stad2939},
	eprint = {2303.02164},
	journal = {Mon. Not. R. Astron. Soc.},
	number = {4},
	pages = {4814--4830},
	primaryclass = {astro-ph.IM},
	title = {{Eryn: a multipurpose sampler for Bayesian inference}},
	volume = {526},
	year = {2023}}

@article{whittlelikelihood,
	author = {P. Whittle},
	issn = {00359246},
	journal = {J. R. Stat. Soc. B},
	number = {1},
	pages = {125--139},
	publisher = {[Royal Statistical Society, Oxford University Press]},
	title = {The Analysis of Multiple Stationary Time Series},
	url = {http://www.jstor.org/stable/2983728},
	urldate = {2024-09-23},
	volume = {15},
	year = {1953}}

@article{Hartwig:2021mzw,
	archiveprefix = {arXiv},
	author = {Hartwig, Olaf and Muratore, Martina},
	doi = {10.1103/PhysRevD.105.062006},
	eprint = {2111.00975},
	journal = {Phys. Rev. D},
	number = {6},
	pages = {062006},
	primaryclass = {gr-qc},
	title = {{Characterization of time delay interferometry combinations for the LISA instrument noise}},
	volume = {105},
	year = {2022}}

@article{Benacquista:2005tm,
	archiveprefix = {arXiv},
	author = {Benacquista, Matthew and Holley-Bockelmann, K.},
	doi = {10.1086/504024},
	eprint = {astro-ph/0504135},
	journal = {Astrophys. J.},
	pages = {589--596},
	title = {{Consequences of disk scale height on LISA confusion noise from close white dwarf binaries}},
	volume = {645},
	year = {2006}}

@article{2025arXiv250812939D,
	adsurl = {https://ui.adsabs.harvard.edu/abs/2025arXiv250812939D},
	archiveprefix = {arXiv},
	author = {{Deistler}, Michael and {Boelts}, Jan and {Steinbach}, Peter and {Moss}, Guy and {Moreau}, Thomas and {Gloeckler}, Manuel and {Rodrigues}, Pedro L.~C. and {Linhart}, Julia and {Lappalainen}, Janne K. and {Miller}, Benjamin Kurt and {Gon{\c{c}}alves}, Pedro J. and {Lueckmann}, Jan-Matthis and {Schr{\"o}der}, Cornelius and {Macke}, Jakob H.},
	eprint = {2508.12939},
	journal = {{}},
	month = aug,
	primaryclass = {stat.ML},
	title = {{Simulation-Based Inference: A Practical Guide}},
	year = 2025}

@article{Green:1995mxx,
	author = {Green, Peter J.},
	doi = {10.1093/biomet/82.4.711},
	journal = {Biometrika},
	number = {4},
	pages = {711--732},
	title = {{Reversible jump Markov chain Monte Carlo computation and Bayesian model determination}},
	volume = {82},
	year = {1995}}

@article{2012ApJ...749L..11B,
	adsurl = {https://ui.adsabs.harvard.edu/abs/2012ApJ...749L..11B},
	archiveprefix = {arXiv},
	author = {{Badenes}, Carles and {Maoz}, Dan},
	doi = {10.1088/2041-8205/749/1/L11},
	eid = {L11},
	eprint = {1202.5472},
	journal = {Astrophys. J. Lett.},
	month = apr,
	number = {1},
	pages = {L11},
	primaryclass = {astro-ph.SR},
	title = {{The Merger Rate of Binary White Dwarfs in the Galactic Disk}},
	volume = {749},
	year = 2012}

@article{Prince:2002hp,
	archiveprefix = {arXiv},
	author = {Prince, Thomas A. and Tinto, Massimo and Larson, Shane L. and Armstrong, J. W.},
	doi = {10.1103/PhysRevD.66.122002},
	eprint = {gr-qc/0209039},
	journal = {Phys. Rev. D},
	pages = {122002},
	title = {{The LISA optimal sensitivity}},
	volume = {66},
	year = {2002}}

@book{2006gpml.book.....R,
	adsurl = {https://ui.adsabs.harvard.edu/abs/2006gpml.book.....R},
	author = {{Rasmussen}, Carl Edward and {Williams}, Christopher K.~I.},
	publisher = {MIT},
	title = {{Gaussian Processes for Machine Learning}},
	year = 2006}

@article{2018MNRAS.480.2704L,
	adsurl = {https://ui.adsabs.harvard.edu/abs/2018MNRAS.480.2704L},
	archiveprefix = {arXiv},
	author = {{Lamberts}, A. and {Garrison-Kimmel}, S. and {Hopkins}, P.~F. and {Quataert}, E. and {Bullock}, J.~S. and {Faucher-Gigu{\`e}re}, C.-A. and {Wetzel}, A. and {Kere{\v{s}}}, D. and {Drango}, K. and {Sanderson}, R.~E.},
	doi = {10.1093/mnras/sty2035},
	eprint = {1801.03099},
	journal = {Mon. Not. R. Astron. Soc.},
	month = oct,
	number = {2},
	pages = {2704-2718},
	primaryclass = {astro-ph.GA},
	title = {{Predicting the binary black hole population of the Milky Way with cosmological simulations}},
	volume = {480},
	year = 2018}

@article{2025A&A...702A.131H,
	adsurl = {https://ui.adsabs.harvard.edu/abs/2025A&A...702A.131H},
	archiveprefix = {arXiv},
	author = {{Hellstr{\"o}m}, L. and {Giersz}, M. and {Askar}, A. and {Hypki}, A. and {Zhao}, Y. and {Lu}, Y. and {Zhang}, S. and {V{\'a}zquez-Aceves}, V. and {Wiktorowicz}, G.},
	doi = {10.1051/0004-6361/202555960},
	eid = {A131},
	eprint = {2506.13122},
	journal = {Astron. Astrophys.},
	month = oct,
	pages = {A131},
	primaryclass = {astro-ph.SR},
	title = {{Formation channels of gravitationally resolvable double white dwarf binaries inside globular clusters}},
	volume = {702},
	year = 2025}

@article{2025A&A...704A.156R,
	adsurl = {https://ui.adsabs.harvard.edu/abs/2025A&A...704A.156R},
	archiveprefix = {arXiv},
	author = {{Rajamuthukumar}, Abinaya Swaruba and {Korol}, Valeriya and {Stegmann}, Jakob and {Preece}, Holly and {Pakmor}, R{\"u}diger and {Justham}, Stephen and {Toonen}, Silvia and {de Mink}, Selma E.},
	doi = {10.1051/0004-6361/202554277},
	eid = {A156},
	eprint = {2502.09607},
	journal = {Astron. Astrophys.},
	month = dec,
	pages = {A156},
	primaryclass = {astro-ph.SR},
	title = {{The role of triple evolution in the formation of LISA double white dwarfs}},
	volume = {704},
	year = 2025}

@article{LISA:2024hlh,
	archiveprefix = {arXiv},
	author = {Colpi, Monica and others},
	eprint = {2402.07571},
	journal = {ESA-SCI-DIR-RP-002},
	month = {2},
	primaryclass = {astro-ph.CO},
	title = {{LISA Definition Study Report}},
	year = {2024}}

@article{Buscicchio:2025zeb,
	archiveprefix = {arXiv},
	author = {Buscicchio, Riccardo and Pozzoli, Federico and Chirico, Daniele and Sesana, Alberto},
	eprint = {2511.03604},
	journal = {{}},
	month = nov,
	primaryclass = {astro-ph.IM},
	title = {{The first year of LISA Galactic foreground}},
	year = {2025}}

@article{Rinaldi:2025evs,
	archiveprefix = {arXiv},
	author = {Rinaldi, Stefano and Toubiana, Alexandre and Gair, Jonathan R.},
	doi = {10.1088/1475-7516/2025/12/031},
	eprint = {2506.05153},
	journal = {J. Cosmology Astropart. Phys.},
	pages = {031},
	primaryclass = {gr-qc},
	title = {{Trust the process: mapping data-driven reconstructions to informed models using stochastic processes}},
	volume = {12},
	year = {2025}}

@article{2001astro.ph..8028P,
	adsurl = {https://ui.adsabs.harvard.edu/abs/2001astro.ph..8028P},
	archiveprefix = {arXiv},
	author = {{Phinney}, E.~S.},
	eprint = {astro-ph/0108028},
	journal = {{}},
	month = aug,
	primaryclass = {astro-ph},
	title = {{A Practical Theorem on Gravitational Wave Backgrounds}},
	year = 2001}

@article{Littenberg:2023xpl,
	archiveprefix = {arXiv},
	author = {Littenberg, Tyson B. and Cornish, Neil J.},
	doi = {10.1103/PhysRevD.107.063004},
	eprint = {2301.03673},
	journal = {Phys. Rev. D},
	number = {6},
	pages = {063004},
	primaryclass = {gr-qc},
	title = {{Prototype global analysis of LISA data with multiple source types}},
	volume = {107},
	year = {2023}}

@article{1977A&A....57..383Z,
	adsurl = {https://ui.adsabs.harvard.edu/abs/1977A&A....57..383Z},
	author = {{Zahn}, J.-P.},
	journal = {Astron. Astrophys.},
	month = may,
	pages = {383-394},
	title = {{Tidal friction in close binary systems.}},
	url = {https://ui.adsabs.harvard.edu/abs/1977A%26A....57..383Z},
	volume = {57},
	year = 1977}

@article{2026arXiv260211765M,
	adsurl = {https://ui.adsabs.harvard.edu/abs/2026arXiv260211765M},
	archiveprefix = {arXiv},
	author = {{McMillan}, Jake and {Ingram}, Adam and {Dashwood Brown}, Cordelia and {Igoshev}, Andrei and {Middleton}, Matthew and {Wiktorowicz}, Grzegorz and {Scaringi}, Simone},
	eprint = {2602.11765},
	journal = {{}},
	month = feb,
	primaryclass = {astro-ph.HE},
	title = {{Population synthesis predictions of the Galactic compact binary gravitational wave foreground detectable by LISA}},
	year = 2026}

@ARTICLE{2025JCAP...06..030K,
       author = {{Kume}, Jun'ya and {Peloso}, Marco and {Pieroni}, Mauro and {Ricciardone}, Angelo},
        title = "{Assessing the impact of unequal noises and foreground modeling on SGWB reconstruction with LISA}",
      journal = {\jcap},
         year = 2025,
        month = jun,
       volume = {2025},
       number = {6},
          eid = {030},
        pages = {030},
          doi = {10.1088/1475-7516/2025/06/030},
archivePrefix = {arXiv},
       eprint = {2410.10342},
 primaryClass = {gr-qc},
       adsurl = {https://ui.adsabs.harvard.edu/abs/2025JCAP...06..030K}
}

\end{document}